\documentclass[12pt]{article}
\catcode`\@=11

\@addtoreset{equation}{section}
\def\theequation{\arabic{equation}}
\def\theequation{\thesection\arabic{equation}}

\newcommand{\be}{\begin{equation}}
\newcommand{\ee}{\end{equation}}
\newcommand{\ba}{\begin{eqnarray}}
\newcommand{\ea}{\end{eqnarray}}

\def\NPB#1#2#3{{\it Nucl.~Phys.} {\bf{B#1}} (19#2) #3}
\def\PLB#1#2#3{{\it Phys.~Lett.} {\bf{B#1}} (19#2) #3}
\def\PRD#1#2#3{{\it Phys.~Rev.} {\bf{D#1}} (19#2) #3}
\def\PRL#1#2#3{{\it Phys.~Rev.~Lett.} {\bf{#1}} (19#2) #3}

\def\JHEP#1#2#3{{\it J. High Energy Phys.} {\bf#1} (19#2) #3}

\def\part{\partial}

\def\e{\epsilon}

\def\h{\eta}

\def\L{\Lambda}

\def\x{\xi}

\def\s{\sigma}

\def\t{\tau}
\def\f{\phi}

\def\w{\omega}
\def\W{\Omega}

\def\@normalsize{\@setsize\normalsize{15pt}\xiipt\@xiipt
\abovedisplayskip 14pt plus3pt minus3pt%
\belowdisplayskip \abovedisplayskip
\abovedisplayshortskip  \z@ plus3pt%
\belowdisplayshortskip  7pt plus3.5pt minus0pt}
\def\small{\@setsize\small{13.6pt}\xipt\@xipt
\abovedisplayskip 13pt plus3pt minus3pt%
\belowdisplayskip \abovedisplayskip
\abovedisplayshortskip  \z@ plus3pt%
\belowdisplayshortskip  7pt plus3.5pt minus0pt
\def\@listi{\parsep 4.5pt plus 2pt minus 1pt
            \itemsep \parsep
            \topsep 9pt plus 3pt minus 3pt}}

\def\underline#1{\relax\ifmmode\@@underline#1\else
        $\@@underline{\hbox{#1}}$\relax\fi}
\@twosidetrue
\relax

\catcode`@=12

\catcode`\@=11

\def\section{\@startsection{section}{1}{\z@}{3.5ex plus 1ex minus
   .2ex}{2.3ex plus .2ex}{\large\bf}}
\def\thesection{\arabic{section}.}
\def\thesubsection{\arabic{section}.\arabic{subsection}}

\def\ps@headings{\def\@oddfoot{}\def\@evenfoot{}
\def\@oddhead{\hbox{}\hfill
        \makebox[.5\textwidth]{\raggedright\ignorespaces --\thepage{}--
        \hfill }}
\def\@evenhead{\@oddhead}
\def\subsectionmark##1{\markboth{##1}{}} }
\renewcommand{\subsection}[1]{\addtocounter{subsection}{1}
\vspace{2.5mm}\par\noindent {\em \thesubsection . #1}\par
 \vspace{0.5mm} }
\ps@headings

\catcode`\@=12

\relax

\def\figcap{\section*{Figure Captions\markboth
        {FIGURECAPTIONS}{FIGURECAPTIONS}}\list
        {Fig. \arabic{enumi}:\hfill}{\settowidth\labelwidth{Fig. 999:}
        \leftmargin\labelwidth
        \advance\leftmargin\labelsep\usecounter{enumi}}}
 \relax
\def\tablecap{\section*{Table Captions\markboth
        {TABLECAPTIONS}{TABLECAPTIONS}}\list
        {Table \arabic{enumi}:\hfill}{\settowidth\labelwidth{Table 999:}
        \leftmargin\labelwidth
        \advance\leftmargin\labelsep\usecounter{enumi}}}
 \relax
\def\reflist{\section*{References\markboth
        {REFLIST}{REFLIST}}\list
        {[\arabic{enumi}]\hfill}{\settowidth\labelwidth{[999]}
        \leftmargin\labelwidth
        \advance\leftmargin\labelsep\usecounter{enumi}}}
 \relax

\catcode`\@=11

\def\marginnote#1{}
\newcount\hour
\newcount\minute
\newtoks\amorpm
\hour=\time\divide\hour by60
\minute=\time{\multiply\hour by60 \global\advance\minute by-
\hour}
\edef\standardtime{{\ifnum\hour<12 \global\amorpm={am}%
    \else\global\amorpm={pm}\advance\hour by-12 \fi
    \ifnum\hour=0 \hour=12 \fi
    \number\hour:\ifnum\minute<100\fi\number\minute\the\amorpm}}
\edef\militarytime{\number\hour:\ifnum\minute<100\fi\number\minute}
\def\draftlabel#1{{\@bsphack\if@filesw {\let\thepage\relax
  \xdef\@gtempa{\write\@auxout{\string
    \newlabel{#1}{{\@currentlabel}{\thepage}}}}}\@gtempa
    \if@nobreak \ifvmode\nobreak\fi\fi\fi\@esphack}
     \gdef\@eqnlabel{#1}}
\def\@eqnlabel{}
\def\@vacuum{}
\def\draftmarginnote#1{\marginpar{\raggedright\scriptsize\tt#1}}
\def\draft{\oddsidemargin -.5truein
        \def\@oddfoot{\sl preliminary draft \hfil
        \rm\thepage\hfil\sl\today\quad\militarytime}
        \let\@evenfoot\@oddfoot \overfullrule 3pt
        \let\label=\draftlabel
        \let\marginnote=\draftmarginnote
   
\def\@eqnnum{(\theequation)\rlap{\kern\marginparsep\tt\@eqnlabel}%
\global\let\@eqnlabel\@vacuum}  }
\def\preprint{\twocolumn\sloppy\flushbottom\parindent 1em
        \leftmargini 2em\leftmarginv .5em\leftmarginvi .5em
        \oddsidemargin -.5in    \evensidemargin -.5in
        \columnsep 15mm \footheight 0pt
        \textwidth 250mmin      \topmargin  -.4in
        \headheight 12pt \topskip .4in
        \textheight 175mm
        \footskip 0pt
        
\def\@oddhead{\thepage\hfil\addtocounter{page}{1}\thepage}
        \let\@evenhead\@oddhead \def\@oddfoot{} \def\@evenfoot{}  }
\def\titlepage{\@restonecolfalse\if@twocolumn\@restonecoltrue\onecolumn
     \else \newpage \fi \thispagestyle{empty}\c@page\z@
        \def\thefootnote{\fnsymbol{footnote}} }
\def\endtitlepage{\if@restonecol\twocolumn \else  \fi
        \def\thefootnote{\arabic{footnote}}
        \setcounter{footnote}{0}}  
\catcode`@=12
\relax

\def\ps@headings{\def\@oddfoot{}\def\@evenfoot{}
\def\@oddhead{\hbox{}\hfill
        \makebox[.5\textwidth]{\raggedright\ignorespaces --\thepage{}--
        \hfill }}
\def\@evenhead{\@oddhead}
\def\subsectionmark##1{\markboth{##1}{}} }

\ps@headings

\relax

\def\firstpage#1#2#3#4#5#6{
\begin{document}


\begin{titlepage}
\nopagebreak
\title{\begin{flushright}
        \vspace*{-1.8in}
        {\normalsize LPTENS 99/38}\\[-10mm]
        {\normalsize CPTH-S743.1099}\\[-10mm]
        {\normalsize NSF-ITP-99-127}\\ [-10mm]
        {\normalsize DFF/347/10/99}\\[-10mm]
        {\normalsize LPT-ORSAY 99/80}\\[-10mm]
        {\normalsize ROM2F-99/40}\\[-10mm]
        {\normalsize hep-th/9911081}\\[-4mm]
\end{flushright}
\vskip 4mm
{#3}}
\vskip -12mm
\author{ #4 \\[0.1cm] #5}
\maketitle
\vskip -9mm     
\nopagebreak 
\begin{abstract} {\noindent #6}
\end{abstract}
\begin{flushleft}
\rule{16.1cm}{0.2mm}\\[-4mm] 
$^{\dagger}${\small Unit{\'e} mixte du CNRS et de l'ENS, UMR 8549.}\\[-4mm]
$^{\ddagger}${\small Unit{\'e} mixte du CNRS et de l'EP, UMR 7644.}\\[-4mm]
$^{\star}${\small Unit{\'e} mixte du CNRS, UMR 8627.}\\
\today
\end{flushleft}
\thispagestyle{empty}
\end{titlepage}}

\date{}
\firstpage{3118}{IC/95/34} {\large\bf Type I vacua with brane supersymmetry
 breaking}  
{C. Angelantonj$^{\,a,b}$, I. Antoniadis$^{\,b,c}$, 
G. D'Appollonio$^{\,d}$,\\[-3mm] E. Dudas$^{\,e}$ and 
A. Sagnotti$^{\,f}$} 
{\small\sl
$^a$ Laboratoire de Physique Th{\'e}orique de l'{\'E}cole Normale 
Sup{\'e}rieure$^\dagger$
\\[-5mm]\small\sl
24, rue Lhomond, F-75231 Paris CEDEX 05
\\[-5mm]\small\sl
$^b$ Centre de Physique Th{\'e}orique$^\ddagger$,  {\'E}cole Polytechnique, 
{}F-91128 Palaiseau\\[-5mm]
\small \sl $^c$ ITP - Kohn Hall, Univ. of California, Santa Barbara,
CA 93106-4030
\\ [-5mm]
\small\sl$^{d}$ Dipartimento di Fisica, Univ. Firenze e INFN, Sez. di
Firenze\\[-5mm]
\small\sl Largo Enrico Fermi 2, I-50125 Firenze \\[-5mm]
\small\sl $^e$  LPT$^\star$, B{\^a}t. 210, Univ. 
Paris-Sud, F-91405 Orsay\\[-5mm] 
\small\sl$^{f}$ Dip. di Fisica, Univ. Roma ``Tor Vergata'' 
e INFN, Sez. di Roma 2\\[-5mm]
\small\sl Via della Ricerca Scientifica 1, I-00133 Roma} 
{We show how chiral type I models whose tadpole conditions have no 
supersymmetric solution can be consistently defined
introducing antibranes with non-supersymmetric world volumes. 
At tree level, the resulting stable non-BPS configurations
correspond to tachyon-free spectra, where supersymmetry is broken at the
string scale on some (anti)branes but is exact in the bulk,  and can be 
further deformed by the addition of brane-antibrane pairs of the same type.
As a result, a scalar potential is generated, that can stabilize some
radii of the compact space.
This setting has the novel virtue of linking supersymmetry breaking to 
the consistency requirements of an underlying fundamental theory.
}


\section{Introduction}

Type I models have become the subject of an intense activity 
during the last few years, since their perturbative definition \cite{carg} 
offers interesting new possibilities for low-energy phenomenology, 
and in particular leaves some freedom to lower the string scale 
well below the Planck mass, if some extra dimensions are 
large \cite{a,w,add,int}. 
Their consistency and a number of their most
amusing features may be traced to the relation to  
suitable ``parent'' models of oriented
closed strings, from which their spectra can be derived. In this 
procedure, a special role is played by ``tadpole conditions'' 
for Ramond-Ramond (RR) states \cite{pc}. These may be regarded as
{\it global} neutrality 
conditions for RR charges \cite{pol}, constrain the
(integer) Chan-Paton multiplicities, and are
usually linked to gauge and gravitational anomalies. 

The explicit study of type I vacua, however, has revealed 
an unexpected difficulty: in some interesting chiral four-dimensional
models it is apparently impossible to satisfy some of the tadpole conditions
\cite{erice,zw}. This peculiar phenomenon can often 
be traced to sign flips of
some crosscap contributions or, in more suggestive space-time language,
to the reversal of some orientifold charges.
Examples are actually known where the solution would lead to negative 
Chan-Paton multiplicities, thus violating the positivity of the
annulus amplitude, or where no solution can be found in general, 
because crosscaps and boundaries scale differently with the
internal volume. 

In a recent work, it was shown how this difficulty can be
evaded, in a prototype six-dimensional $Z_2$ model and in the 
four-dimensional $Z_2 \times Z_2$ orientifold, if one relaxes the condition
that the brane configuration be supersymmetric, thus allowing for vacua
including both branes and antibranes. As a result,
supersymmetry is broken on some collection of branes
{\it at the string scale}, 
while it is preserved (to lowest order) in the bulk and
possibly on other branes. Aside from its interest for the consistent definition
of type I models, this scenario, termed in \cite{ads2} ``brane
supersymmetry breaking'',  has clearly some  beauty of its own if our
non-supersymmetric universe is modeled as a brane in a bath of
higher-dimensional supergravity.  Rather than being introduced as a possible
deformation, the breaking of  supersymmetry in our low-energy world would then
be seen as a neat consequence of the internal consistency of the 
underlying String Theory. Moreover, in this context 
the present experimental limits on small-distance deviations from 
Newtonian gravity leave open the exciting possibility that supersymmetry, 
broken on our world brane, be almost exact a millimeter away from it.
We would like to stress that this mechanism links supersymmetry
breaking to the string scale, rather than to geometric scales of the
internal space, as Scherk-Schwarz \cite{closed,ads,aaf,adds2} or 
magnetic \cite{bachas} deformations. 
The natural distinction between these 
two settings, however, is closely linked to 
a geometric interpretation of the two-dimensional Conformal Field Theory, 
and is somewhat blurred in non-geometric 
settings, such as those recently explored in \cite{bmp}.

Actually, all these vacua can be further deformed. In particular,
in type I models one can add arbitrary numbers of D9 -- D$\bar 9$ and
D5 -- D$\bar 5$ brane pairs
without violating any RR tadpole condition \cite{su,au}, but introducing
additional breakings of supersymmetry on the branes. 
For instance, starting from the $Z_3$ orientifold of \cite{abpss}, one
can thus build semi-realistic three-generation models \cite{aiq}. 
We would like to stress that the introduction of
(anti)5-branes requires in general that additional $Z_2$ projections 
act on the boundaries consistently with the closed spectrum. 
In this case the
resulting configurations are generally unstable, due to the presence 
of tachyonic modes that, however, can often be lifted 
by suitable Wilson lines (or, equivalently, by suitable brane displacements). 
This should be contrasted with the original setting
for brane supersymmetry breaking \cite{ads2}, where the brane 
configuration involves branes
and antibranes of different types and is by construction stable. Still, 
in both settings the models include non-BPS (anti)brane systems of the
type considered in \cite{sen}, whose interactions with gravity are
consistently described by type I string theory.

In this work we discuss in more detail the open descendants
of two typical examples of type IIB orbifolds, where the problem of unsolvable
tadpoles was first encountered. We begin in Section 2 with a detailed
discussion of  four-dimensional $Z_2 \times Z_2$ models with discrete torsion,
showing how the unavoidable reversal of some of the orientifold charges is
naturally accompanied, in a fully consistent construction, by the
simultaneous presence of branes and antibranes. 
We also discuss similar modifications of the open descendants of the
corresponding freely-acting orbifolds recently studied in
\cite{adds2}. In Section 3 we turn to the $T^6/Z_4$ orbifold, a canonical 
case where the
tadpole conditions can not be solved. Here we can display a
consistent non-supersymmetric solution, provided the Klein-bottle 
projection is also
modified, with the net result of lifting the offending twisted tadpole. 
We should stress, however, that the difficulties met in these two
classes of orbifolds are rather distinct. In type-I models,
the twisted tadpoles are generally related to
non-abelian gauge anomalies,
whereas the untwisted ones are related to gravitational anomalies, 
absent in four dimensions. In the $Z_2 \times Z_2$ models
with discrete torsion the
twisted tadpoles are actually {\it solvable}, whereas the untwisted ones
are not, and as a result even the naive supersymmetric open spectrum is
free of non-abelian gauge anomalies \cite{erice}.
On the other hand, in the $Z_4$ model with standard Klein bottle
the problem is related to the twisted tadpoles, and as a result the
naive supersymmetric open spectrum has non-abelian gauge anomalies \cite{zw}.
In Section 4 we discuss the possible deformations of type I models to stable
vacuum configurations including both branes and antibranes of the same type.
In particular, we discuss deformations of
six-dimensional toroidal compactifications, that may thus
include arbitrary pairs
of D9 -- D$\bar 9$ and D5 -- D$\bar 5$ branes, and present a similar
generalization of the four-dimensional $T^6/Z_4$ orientifold studied 
in Section 3.
We show explicitly how the presence of additional brane-antibrane pairs,
via the resulting NS-NS (Neveu-Schwarz) tadpoles, 
can actually stabilize the radii of the 
compact internal space. All the models presented in this paper are meant
to provide new instances of the phenomenon discussed in \cite{ads2}:
to lowest order, the D-branes
are supersymmetric (with suitable diagonal subgroups of the antibrane 
gauge groups realized as 
global symmetries), while the antibranes are not. 
Finally, Section 5 contains our Conclusions and the two Appendices collect
some relevant properties of the characters used in the text. 


\section{$Z_2\times Z_2$ orientifolds and discrete torsion}

Let us begin by displaying the torus amplitude for the parent type IIB 
$Z_2 \times Z_2$ orbifold. Aside from the 
identity, that we denote by $o$, the other three elements act on the three 
internal two-tori as 
\be
g: (+,-,-) \quad , \qquad f: (-,+,-) \quad , \qquad h: (-,-,+) \quad .
\label{a4}
\ee
In the $Z_2 \times Z_2$ orbifold, one has the freedom of introducing discrete 
torsion \cite{vafa}, that corresponds to associating a sign 
$\e = \pm 1$ to the 
independent modular orbit containing all  
terms that are twisted and projected by two different orbifold operations.
Omitting for brevity the contributions of the space-time bosons,
the torus amplitudes of the two $Z_2 \times Z_2$ models are 
\ba
{\cal T}\!\!\!&=&\!\!{1 \over 4} \Biggl\{ 
|T_{oo}|^2 \Lambda_1 \Lambda_2 \Lambda_3
+|T_{og}|^2  \Lambda_1 
\left|{4\eta^2 \over \vartheta_2^2}\right|^2 + |T_{of}|^2 \Lambda_2 
\left|{4\eta^2 \over \vartheta_2^2}\right|^2+
|T_{oh}|^2 \Lambda_3 
\left|{4\eta^2 \over \vartheta_2^2}\right|^2 
\nonumber \\
&+& |T_{go}|^2 \Lambda_1
\left|{4\eta^2 \over \vartheta_4^2}\right|^2+ |T_{gg}|^2  \Lambda_1 
\left|{4\eta^2 \over \vartheta_3^2}\right|^2 
+ |T_{fo}|^2 
\Lambda_2 \left|{4 \eta^2 \over \vartheta_4^2}\right|^2 + 
|T_{ff}|^2 \Lambda_2 \left|{4 \eta^2 \over \vartheta_3^2}\right|^2  
\nonumber \\
&+& |T_{ho}|^2 \Lambda_3
\left|{4\eta^2 \over \vartheta_4^2}\right|^2 +|T_{hh}|^2 
\Lambda_3 \left|{4\eta^2 \over \vartheta_3^2}\right|^2 
\nonumber \\
&+& \epsilon \left( |T_{gh}|^2 + |T_{gf}|^2 + |T_{fg}|^2 + |T_{fh}|^2 
+ |T_{hg}|^2
+ |T_{hf}|^2 \right) \left|{8\eta^3 \over \vartheta_2 \vartheta_3 
\vartheta_4} \right|^2 
\Biggr\} \quad  , 
\label{a1} 
\ea
where the $\L_k$ are lattice sums for the three internal 
tori and, throughout the paper, we let $\alpha'=2$. We have 
expressed the torus amplitude in terms of the 16 
quantities $(k=o,g,h,f)$
\ba
T_{ko} &=&  \tau_{ko} +  \tau_{kg} + \tau_{kh} + \tau_{kf} \quad , \qquad
T_{kg} =  \tau_{ko} +  \tau_{kg} - \tau_{kh} - \tau_{kf} \quad , \nonumber \\
T_{kh} &=&  \tau_{ko} -  \tau_{kg} + \tau_{kh} - \tau_{kf} \quad , \qquad
T_{kf} =  \tau_{ko} -  \tau_{kg} - \tau_{kh} + \tau_{kf} \quad ,
\label{a2}
\ea
where the $16$ $Z_2 \times Z_2$ characters $\tau_{kl}$ \cite{erice}, 
combinations of products of level-one SO(2) characters, are displayed
in Appendix A.
The choices $\e = \mp 1$ identify the models with and 
without discrete torsion, whose low-energy spectra are quite different: in 
the first ${\cal N}=2$ supergravity is coupled to 52 hypermultiplets and 
3 vector multiplets, while in the second it is coupled to 4 
hypermultiplets and 51 vector 
multiplets. These spectra correspond to orbifold limits of Calabi-Yau 
manifolds with Hodge numbers $(51,3)$ and $(3,51)$, respectively. 


The $\W$ projections that we are considering are implemented by the 
Klein-bottle amplitudes
\ba
{\cal K} &=&\frac{1}{8} \biggl\{ ( P_1 P_2 P_3 + P_1 W_2 W_3 + W_1 P_2
W_3 + W_1 W_2 P_3 ) T_{oo} \nonumber \\ &+ & \!\! 
2 \times 16 \bigl[\epsilon_1 (P_1 + \epsilon W_1 ) T_{go} 
+  \epsilon_2 (P_2 + \epsilon W_2 ) T_{fo}
+ \epsilon_3 (P_3 + \epsilon W_3 ) T_{ho} \bigr] 
\left( \frac{\eta}{\vartheta_4} \right)^2 \biggr\} \ ,
\label{a5}
\ea
where $P_k$ and $W_k$ denote the restrictions of the lattice sums $\L_k$ to 
their momentum and winding sublattices.
Discrete torsion has a neat effect \cite{adds2} 
on $(P_k+\e W_k)$: 
if $\e= 1$, the massless 
twisted contributions are diagonal combinations of the $\t_{kl}$, and
appear in the Klein bottle, while  
if $\e = -1$ they are off-diagonal 
combinations, and do not contribute to it.  Consistently with 
the crosscap constraint \cite{pss}, ($\ref{a5}$) can accommodate 
three additional 
signs $\e_k$. Actually, these are not independent, but are linked to 
the parameter $\e$ by the constraint
\be
\epsilon_1 \ \epsilon_2 \ \epsilon_3 \ = \epsilon \quad .
\label{a6}
\ee
One can write this amplitude as
\be
{\cal K} = \frac{1}{8} \biggl\{ (P_1 P_2 P_3 + \frac{1}{2} P_k W_l W_m) T_{oo}
+ 2 \times 16 \epsilon_k (P_k + \epsilon W_k ) T_{ko}
\left( \frac{\eta}{\vartheta_4} \right)^2  
\biggr\} \quad ,
\label{a7}
\ee
where we have resorted to a compact notation, used extensively
in the following: {\it summations} over 
repeated indices and {\it symmetrizations} over 
distinct indices are left implicit. An $S$ 
transformation turns this expression into the 
corresponding vacuum-channel amplitude
\be
\tilde{\cal K} = \frac{2^5}{8} \biggl\{ \bigl( v_1v_2v_3 W^e_1 W^e_2 
W^e_3 + \frac{v_k}{2v_l v_m} W_k^e P_l^e P_m^e  \bigr) T_{oo} 
+ 2 \epsilon_k (v_k W_k^e + \epsilon \frac{P_k^e}{v_k}) T_{ok}
\left( \frac{2\eta}{\vartheta_2} \right)^2 \biggr\} \ ,
\label{a8}
\ee
where the superscript $e$ denotes the restriction of the
lattice sums to their even terms and the $v_k$ denote the volumes of the
three internal tori. At the origin of the lattices, the constraint
(\ref{a6}) leads to an expression whose coefficients are perfect squares,
\ba
\tilde{\cal K}_0 &=& \frac{2^5}{8} \Biggl\{ \ \left( \sqrt{v_1v_2v_3} +
 \epsilon_1 \sqrt{\frac{v_1}{v_2 v_3}} + 
\epsilon_2 \sqrt{\frac{v_2}{v_1 v_3}}  + \epsilon_3
\sqrt{\frac{v_3}{v_1 v_2}} \right)^2 \tau_{oo} 
\nonumber \\ 
& &+  \left( \sqrt{v_1v_2v_3} +
 \epsilon_1 \sqrt{\frac{v_1}{v_2 v_3}} - 
\epsilon_2 \sqrt{\frac{v_2}{v_1 v_3}}  - \epsilon_3
\sqrt{\frac{v_3}{v_1 v_2}} \right)^2 \tau_{og} 
\nonumber \\
& &+ \left( \sqrt{v_1v_2v_3} -
 \epsilon_1 \sqrt{\frac{v_1}{v_2 v_3}} + 
\epsilon_2 \sqrt{\frac{v_2}{v_1 v_3}}  - \epsilon_3
\sqrt{\frac{v_3}{v_1 v_2}} \right)^2 \tau_{of} 
\nonumber \\
&&+ \left( \sqrt{v_1v_2v_3} -
 \epsilon_1 \sqrt{\frac{v_1}{v_2 v_3}} - 
\epsilon_2 \sqrt{\frac{v_2}{v_1 v_3}}  + \epsilon_3
\sqrt{\frac{v_3}{v_1 v_2}} \right)^2 \tau_{oh} 
 \ \Biggr\} \quad ,
\label{a9}
\ea
that shows rather neatly how
the choice $\e_k=-1$ {\it reverts} the charge of the O$5_k$ orientifold
plane. While
manifestly compatible with the usual positivity requirements, this reversal
clearly affects the tadpole conditions, that
require the introduction of antibranes. In this respect, it should be
appreciated that, according to (\ref{a6}), discrete torsion implies the
reversal of at least one of the O5 charges.
Therefore, taking into account the presence of the $\e_k$, one can identify
four classes of models, determined by the independent choices for
$(\e_1, \e_2, \e_3)$.
If $\e=1$, the choice $(+,+,+)$ gives the model discussed in \cite{erice,bl},
with $48$ chiral 
multiplets from the closed twisted sectors, while the choice $(+,-,-)$ 
gives a model with $16$ chiral multiplets and $32$ vector multiplets from the 
twisted sectors. 
On the other hand, for $\e = -1$ the two choices $(+,+,-)$ 
and $(-,-,-)$ yield the same massless twisted spectrum, 
namely $48$ chiral multiplets. 


In order to describe the annulus amplitude, it is 
convenient to introduce a compact notation. If
$T_{kl}^{\rm NS}$ ($T_{kl}^{\rm R}$) denote the NS (R) parts of the usual 
combinations of  supersymmetric $Z_2 \times Z_2$ characters,
we thus define
\be
\tilde{T}_{kl}^{(\varepsilon)} = T_{kl}^{\rm NS} - \varepsilon 
T_{kl}^{\rm R} \quad ,
\label{a10}
\ee
where $\varepsilon = \pm 1$. Whereas
$\tilde{T}_{kl}^{(+)}(={T}_{kl}^{(+)})$, that in the following we
simply denote by $T_{kl}$ for the sake of brevity, form a closed set under
$S$ modular transformations, the additional combinations
$\tilde{T}_{kl}^{(-)}$ with reversed RR charges, associated to 
the interactions between branes and antibranes, do not.
As a result, the corresponding terms in ${\cal A}$, describing open strings 
stretched between branes and antibranes, contain new combinations
$T^{(-)}_{kl}$, obtained from the $T^{(+)}_{kl}$ 
interchanging $O_2$ with $V_2$ and $S_2$ with $C_2$ in the last 
three factors, 
as explained in Appendix A. 

The transverse-channel annulus amplitude is 
\ba
\tilde{{\cal A}} \!\!\!&=&\!\!\!\frac{2^{-5}}{8} \Biggl\{
\left( N_o^2 v_1v_2v_3W_1W_2W_3 + \frac{D_{k;o}^2v_k}{2v_lv_m} W_k P_l P_m 
\right) T_{oo} 
\nonumber 
\\
&+& 4\left[ ( N_k^2 + D^2_{k;k} ) v_kW_k  + D^2_{l \neq k;k} 
\frac{P_k}{v_k}\right] 
T_{ko} \left( \frac{2\eta}{\vartheta_4} \right)^2 
\\
\!\!\!&+&\!\!\! 2 N_o D_{k;o} v_kW_k {\tilde T}_{ok}^{( \e_k )}
\left( \frac{2\eta}{\vartheta_2} \right)^2 
+ 2 N_k D_{k;k} v_kW_k {\tilde T}_{kk}^{(\e_k)} 
\left( \frac{2\eta}{\vartheta_3} \right)^2 
+ 4 N_l D_{k \neq l;l}{\tilde T}_{lk}^{(\e_k)}
\frac{8 \eta^3}{\vartheta_2 \vartheta_3 \vartheta_4} 
\nonumber \\ 
\!\!\!\!&+&\!\!\!\! 
D_{k;o} D_{l;o} \frac{P_m}{v_m} {\tilde T}_{om}^{(\e_k \e_l )}
\left( \frac{2\eta}{\vartheta_2} \right)^2  
\!\!+ D_{k;m}D_{l;m} \frac{P_m}{v_m} {\tilde T}_{mm} ^{(\e_k \e_l )} 
\left( \frac{2\eta}{\vartheta_3} \right)^2 
\!\!+ 4 D_{k;k}D_{l;k} {\tilde T}_{km} ^{(\e_k \e_l )}
\frac{8 \eta^3}{\vartheta_2 \vartheta_3 \vartheta_4} 
 \Biggr\} \nonumber \quad ,
\label{a14}
\ea
where $N_o$, $D_{g;o}$, $D_{f;o}$ and
$D_{h;o}$ are the charges for the D9 branes and for the three 
sets of D5 or D${\bar 5}$ branes wrapped around the first, second and 
third torus. In a similar fashion, $N_k$, $D_{g;k}$, $D_{f;k}$ and $D_{h;k}$ 
$(k=g,f,h)$ parametrize the breakings induced by the three orbifold 
operations $g$, $f$ and $h$.
As expected, the RR part of every term describing the interaction between
a brane and an antibrane has a reversed sign.
The untwisted terms at the origin of the lattice sums rearrange 
themselves into perfect squares:
\ba
\tilde{\cal A}_0 &=&
\frac{2^{-5}}{8} \Biggl\{ \ \left( N_o \sqrt{v_1v_2v_3} +
 D_{g;o} \sqrt{\frac{v_1}{v_2 v_3}} + 
D_{f;o} \sqrt{\frac{v_2}{v_1 v_3}}  + D_{h;o}
\sqrt{\frac{v_3}{v_1 v_2}} \right)^2 \tau^{\rm NS}_{oo} 
\nonumber \\ 
& &- \left( N_o \sqrt{v_1v_2v_3} +
 \epsilon_1 D_{g;o} \sqrt{\frac{v_1}{v_2 v_3}} + 
\epsilon_2 D_{f;o} \sqrt{\frac{v_2}{v_1 v_3}}  + \epsilon_3
D_{h,o} \sqrt{\frac{v_3}{v_1 v_2}} \right)^2 \tau^{\rm R}_{oo} 
\nonumber\\
& &+  \left( N_o \sqrt{v_1v_2v_3} +
 D_{g;o} \sqrt{\frac{v_1}{v_2 v_3}} - 
D_{f;o} \sqrt{\frac{v_2}{v_1 v_3}}  - D_{h;o}
\sqrt{\frac{v_3}{v_1 v_2}} \right)^2 \tau^{\rm NS}_{og} 
\nonumber \\
& &- \left( N_o\sqrt{v_1v_2v_3} +
\epsilon_1 D_{g;o} \sqrt{\frac{v_1}{v_2 v_3}} -
\epsilon_2 D_{f;o} \sqrt{\frac{v_2}{v_1 v_3}}  - \epsilon_3
D_{h;o} \sqrt{\frac{v_3}{v_1 v_2}} \right)^2 \tau^{\rm R}_{og} 
\nonumber\\
&&+ \left( N_o \sqrt{v_1v_2v_3} -
 D_{g;o} \sqrt{\frac{v_1}{v_2 v_3}} + 
D_{f;o} \sqrt{\frac{v_2}{v_1 v_3}}  - D_{h;o}
\sqrt{\frac{v_3}{v_1 v_2}} \right)^2 \tau^{\rm NS}_{of} 
\nonumber \\ 
& &- \left( N_o \sqrt{v_1v_2v_3} -
 \epsilon_1 D_{g;o} \sqrt{\frac{v_1}{v_2 v_3}} + 
\epsilon_2 D_{f;o} \sqrt{\frac{v_2}{v_1 v_3}}  - \epsilon_3
D_{h;o} \sqrt{\frac{v_3}{v_1 v_2}} \right)^2 \tau^{\rm R}_{of} 
\nonumber\\
& &+ \left( N_o \sqrt{v_1v_2v_3} -
 D_{g;o} \sqrt{\frac{v_1}{v_2 v_3}} - 
D_{f;o} \sqrt{\frac{v_2}{v_1 v_3}}  + D_{h;o}
\sqrt{\frac{v_3}{v_1 v_2}} \right)^2 \tau^{\rm NS}_{oh} 
\nonumber \\
& &- \left( N_o\sqrt{v_1v_2v_3} -
 \epsilon_1 D_{g;o} \sqrt{\frac{v_1}{v_2 v_3}} - 
\epsilon_2 D_{f;o} \sqrt{\frac{v_2}{v_1 v_3}}  + \epsilon_3
D_{h;o} \sqrt{\frac{v_3}{v_1 v_2}} \right)^2 \tau^{\rm R}_{oh} 
 \ \biggr\} \, .
\label{a15}
\ea
Moreover, the breaking terms reflect rather neatly the 
geometry of the (anti)brane configuration. Indeed, the coefficient 
that multiplies a 
given twisted character is a sum of squares associated to 
the fixed tori of the various twisted sectors, and each square contains 
the breaking terms
for the branes present in the fixed tori, with factors $\sqrt{v}$ if 
they are wrapped around them or $1/\sqrt{v}$ if they are localized 
on them. The relative coefficients of these terms are also directly linked
to the brane geometry, and are given by
\be
\sqrt{\frac{\rm \# \ of \ fixed \ tori}{\rm \# \ of 
\ occupied \ fixed \ tori}} 
\quad  .
\label{a15b}
\ee
Thus, for a given twisted sector, the numerator counts the fixed 
tori, while the denominator counts the fixed tori where branes
are actually present.  Moreover, the R portions of the characters 
describing brane-antibrane exchanges have reverted signs also in
these twisted contributions, as expected. 
For instance, in the 
$g$-twisted sector of the $(++-)$ model, that  contains D${\bar 5}_3$
branes, the reflection coefficients for the massless modes in $\tau_{gh}$ are
\ba
& & \frac{2^{-5}}{8} \biggl[\left( N_g \sqrt{v_1}-4D_{g;g}\sqrt{v_1}
-2D_{f;g}\frac{1}{\sqrt{v_1}}+2D_{h;g}\frac{1}{\sqrt{v_1}} \right)^2 
+3 \left( N_g\sqrt{v_1}-2D_{f;g}\frac{1}{\sqrt{v_1}} \right)^2 
\nonumber \\
& &\qquad + 3\left( N_g\sqrt{v_1}+2D_{h;g}\frac{1}{\sqrt{v_1}} \right)^2 
+9N_g^2v_1 \ \biggr] 
\label{a15c}
\ea
for the NS portion, and
\ba
& & \frac{2^{-5}}{8} \biggl[\left(N_g\sqrt{v_1}-4D_{g;g}\sqrt{v_1}
-2D_{f;g}\frac{1}{\sqrt{v_1}}-2D_{h;g}\frac{1}{\sqrt{v_1}} \right)^2  
+3 \left( N_g\sqrt{v_1}-2D_{f;g}\frac{1}{\sqrt{v_1}} \right)^2 
\nonumber \\
& & \qquad + 3 \left( N_g\sqrt{v_1}-2D_{h;g}\frac{1}{\sqrt{v_1}} \right)^2 
+9N_g^2v_1 \ \biggr] 
\label{a15d}
\ea
for the R portion. According to 
($\ref{a15b}$),
the coefficient of $N_g$ is $\sqrt{v_1}$, since the D9 are wrapped 
around all fixed tori, the coefficient of $D_{g;g}$ is $4\sqrt{v_1}$, 
since the D$5_1$ are only wrapped 
around one fixed torus, while the coefficients of $D_{f;g}$ and  
$D_{h;g}$ are $2/\sqrt{v_1}$, since the D$5_2$ and D${\bar 5}_3$ are 
confined to four of the fixed tori. Finally, out of the 16 $g$-fixed tori, 
one sees all the branes, three see only the 
D9 and the D$5_2$, three see only the D9 and the D${\bar 5}_3$ and,
finally,  nine see only the D9.

The direct-channel annulus amplitude is then
\ba
{\cal A} &=& \frac{1}{8} \Biggl\{
\left( N_o^2 P_1P_2P_3 + \frac{D_{k;o}^2}{2} P_k W_l W_m \right) T_{oo} + 
\left[ ( N_k^2 + D^2_{k;k} ) P_k  + D^2_{l \neq k ;k} W_k\right] T_{ok}
\left( \frac{2\eta}{\vartheta_2} \right)^2 
\nonumber
\\
&+& 2 N_o D_{k;o} P_k T_{ko}^{( \e_k )}
\left( \frac{\eta}{\vartheta_4} \right)^2 
- 2 N_k D_{k;k} P_k T_{kk}^{(\e_k)}
\left( \frac{\eta}{\vartheta_3} \right)^2  
\\
&+& 2 i (-1)^{k+l} N_l D_{k \neq l ;l}T_{kl}^{(\e_k)}
\frac{2\eta^3}{\vartheta_2 \vartheta_3 \vartheta_4}
+ D_{k;o} D_{l;o} W_m  T_{mo}^{(\e_k \e_l )}
\left( \frac{\eta}{\vartheta_4} \right)^2   
\nonumber
\\
&-& D_{k;m}D_{l;m} W_m T_{mm} ^{(\e_k \e_l )}
\left( \frac{\eta}{\vartheta_3} \right)^2   
+ 2 i (-1)^{m+k} D_{k;k}D_{l;k} T_{mk}^{(\epsilon_k \epsilon_l)} 
\frac{2\eta^3}{\vartheta_2 \vartheta_3 \vartheta_4} 
\Biggr\} \ , \nonumber 
\label{a13bis}
\ea
where in the signs $(-1)^{k+l}$ and $(-1)^{m+k}$ 
$k,l,m$ take the values $1,2,3$ for the $g$, $f$, and $h$ generators.
The transverse-channel amplitudes $\tilde{\cal K}$ and  $\tilde{\cal A}$
determine by standard methods the transverse M{\"o}bius 
amplitude  
\ba
\tilde{{\cal M}} &=& 
- \frac{1}{4} \Biggl\{ N_o v_1v_2v_3W^e_1 W^e_2 W^e_3 \hat{T}_{oo} 
+ N_o v_kW^e_k \epsilon_k\hat{T}_{ok}
\left( \frac{2 \hat{\eta}}{\hat{\vartheta_2}}\right)^2  
+ \frac{v_k}{2v_lv_m}D_{k;o} W^e_k P^e_l P^e_m  \epsilon_k 
\hat{\tilde T}{}_{oo} ^{(\e_k)} 
\nonumber 
\\ 
&+& \left( D_{l;o} \epsilon_k \frac{P^e_m}{v_m} 
\hat{\tilde T}{}_{om}^{(\epsilon_l)} 
+ D_{k;o} v_kW^e_k \hat{\tilde T}{}_{ok}^{(\e_k)} \right)
\left( \frac{2 \hat{\eta}}{\hat{\vartheta_2}}\right)^2  \Biggr\} 
\label{a16}
\ea
and, after a $P$ transformation, the direct-channel M{\"o}bius
amplitude
\ba
{\cal M} &=& - \frac{1}{8} \biggl\{ N_o P_1 P_2 P_3 \hat{T}_{oo} 
- N_o P_k \epsilon_k
\hat{T}_{ok} \left( \frac{2 \hat{\eta}}{\hat{\vartheta_2}}\right)^2 
+ \frac{1}{2}D_{k;o} P_k W_l W_m  \epsilon_k \hat{\tilde T}{}_{oo} ^{(\e_k)} 
\nonumber 
\\ 
&-& \left(D_{l;o} \epsilon_k W_m \hat{\tilde T}{}_{om}^{(\epsilon_l)} 
+ D_{k;o} P_k \hat{\tilde T}{}_{ok}^{(\e_k)}\right)
\left( \frac{2 \hat{\eta}}{\hat{\vartheta_2}}\right)^2 
\biggr\} \quad .
\label{a17}
\ea

From the transverse amplitudes one can now read the tadpole conditions
\ba
& N_o = 32 \ , \qquad & N_g = N_f = N_h = 0 \ , 
\nonumber 
\\
& D_{k;o} = 32 \ , \qquad & D_{k;g} = D_{k;f} = D_{k;h} = 0 \ .
\ea

\subsection{Massless Spectra}


The models where only one $\e_k$ is negative have discrete torsion 
and contain one D${\bar 5}$. For the D9 and the 
two sets of D5 branes, the gauge groups are ${\rm U}(8)\times {\rm U}(8)$, 
with ${\cal N}=1$ supersymmetry, while for the 
D${\bar 5}$ branes the gauge group is ${\rm USp}(8)^4$, with ${\cal N}=0$. 
Moreover, the 59 and 
$5_k5_l$ strings are supersymmetric, while the $9\bar{5}$ and
$5_k\bar{5}$ strings are not. 
Let us discuss in some detail the case $(\e_1,\e_2,\e_3) = (++-)$, that 
contains 
D${\bar 5}$ branes wrapped around the third torus. To this end, let us
parametrize the charges as
\ba
N_o &=& o+g+\bar{o}+\bar{g} \ , \hspace{2.4cm} 
N_g \ \ = \ \ i(o+g-\bar{o}-\bar{g}) \ ,
\nonumber \\
N_f &=& i(o-g-\bar{o}+\bar{g}) \ , \hspace{1.9cm}
N_h \ \ = \ \ o-g+\bar{o}-\bar{g} \ ,
\nonumber \\
D_{g;o} &=& o_1+g_1+\bar{o}_1+\bar{g}_1 \ , \hspace{1.5cm}  
D_{g;g}  \ \ =  \ \ i(o_1+g_1-\bar{o}_1-\bar{g}_1) \ ,\nonumber \\
D_{g;f} &=& o_1-g_1+\bar{o}_1-\bar{g}_1 \ , \hspace{1.5cm} 
D_{g;h}  \ \ = \ \  -i(o_1-g_1-\bar{o}_1+\bar{g}_1) \ ,
\nonumber \\
D_{f;o} &=& o_2+g_2+\bar{o}_2+\bar{g}_2 \ , \hspace{1.5cm} 
D_{f;g} \ \ = \ \ o_2-g_2+\bar{o}_2-\bar{g}_2 \ ,\nonumber \\
D_{f;f} &=& i(o_2+g_2-\bar{o}_2-\bar{g}_2) \ , \hspace{1cm} 
D_{f;h}  \ \ =  \ \ i(o_2-g_2-\bar{o}_2+\bar{g}_2) \ ,
\nonumber \\
D_{h;o} &=& a+b+c+d \ , \hspace{2.2cm} 
D_{h;g}  \ \ =  \ \ a+b-c-d \ ,\nonumber \\
D_{h;f} &=& a-b+c-d \ , \hspace{2.2cm} 
D_{h;h}  \ \ =  \ \ a-b-c+d \ ,
\label{a19}
\ea
and extract the massless spectrum from the amplitudes at the 
origin of the lattices.
The $99$, $5_15_1$ and $5_25_2$ sectors have ${\cal N}=1$ supersymmetry, 
and all give gauge groups ${\rm U}(8) \times {\rm U}(8)$, with chiral 
multiplets in the representations
$(8,8)$, $(8,\bar{8})$, $(28,1)$, $(1,28)$ 
and their conjugates. Moreover, as expected, the $95_1$, $95_2$ and $5_15_2$ 
strings are also supersymmetric, and contain chiral multiplets in the 
representations 
\ba
95_1 : \ & & (8,1;1,\bar{8}) \ , \ \ (1,8;\bar{8},1) \ , 
          \ \ (\bar{8},1;8,1) \ , \ \ (1,\bar{8};1,8) \ , 
\nonumber \\
95_2 : \ & & (8,1;1,\bar 8) \ , \ \ (1,\bar 8;\bar 8,1) \ , 
          \ \ (\bar 8,1;8,1) \ , \ \ (1,8;1,8) \ , 
\nonumber \\
5_15_2 : \  & &  (8,1;8,1) \ , \ \ (1,8;\bar{8},1) \ , 
          \ \ (\bar{8},1;1,8) \ , \ \ (1,\bar{8};1,\bar{8}) \ .
\nonumber
\ea 

On the other hand, the strings whose ends live on the antibrane give rise
to supersymmetric spectra, even if the annulus contains supersymmetric 
characters, since bosons and fermions are treated differently by ${\cal M}$. 
Thus, the $\bar{5}_3\bar{5}_3$ sector contributes a gauge group 
${\rm USp}(8)^4$, with Weyl spinors in the $(28,1,1,1)$ and in three
additional permutations, and chiral multiplets in the 
$(8,8,1,1)$ and in five additional permutations. Finally,
the strings stretched between a brane
and an antibrane have non-supersymmetric spectra, with
Weyl spinors and complex scalars in the representations
\ba
9\bar{5}_3  &{\rm spinors \ :}& \
  (\bar{8},1;8,1,1,1) \ , \ \ (1,\bar{8};1,8,1,1) \ , 
          \ \ (1,8;1,1,8,1) \ , \ \ (8,1;1,1,1,8) \ , 
\nonumber \\
           &{\rm scalars \ :}&  \ 
(\bar{8},1;1,8,1,1) \ , \ \ (1,\bar{8};8,1,1,1) \ , 
          \ \ (1,8;1,1,1,8) \ , \ \ (8,1;1,1,8,1) \ , 
\nonumber \\
5_1\bar{5}_3 &{\rm spinors \ :}& \
  (\bar{8},1;1,1,8,1) \ , \ \ (1,\bar{8};1,1,1,8) \ , 
          \ \ (1,8;1,8,1,1) \ , \ \ (8,1;8,1,1,1) \ , 
\nonumber \\
           &{\rm scalars \ :}&  \ 
(\bar{8},1;1,1,1,8) \ , \ \ (1,\bar{8};1,1,8,1) \ , 
         \ \ (1,8;8,1,1,1) \ , \ \ (8,1;1,8,1,1) \ , 
\nonumber \\
5_2\bar{5}_3 &{\rm spinors \ :}& \
  (8,1;8,1,1,1) \ , \ \ (1,\bar{8};1,1,1,8) \ , 
          \ \ (1,8;1,1,8,1) \ , \ \ (\bar{8},1;1,8,1,1) \ , 
\nonumber \\
           &{\rm scalars \ :}&  \ 
(\bar{8},1;1,1,1,8) \ , \ \ (1,\bar{8};1,8,1,1) \ , 
          \ \ (1,8;8,1,1,1) \ , \ \ (8,1;1,1,8,1) \ . \nonumber
\ea
The choice $(\e_1,\e_2,\e_3) = (---)$ also corresponds to a model with 
discrete torsion. In this case, however, there are D9 branes and three sets of 
D${\bar 5}$ branes, while the charges are to be parametrized as
\ba
N_o &=& a+b+c+d \ , \hspace{2.4cm} 
N_g  \ \ =  \ \ a+b-c-d \ ,\nonumber \\
N_f &=& a-b+c-d \ , \hspace{2.4cm} 
N_h  \ \ =  \ \ a-b-c+d \ , \nonumber \\
D_{g;o} &=& o_1+g_1+\bar{o}_1+\bar{g}_1 \ , \hspace{1.5cm} 
D_{g;g}  \ \ =  \ \ o_1-g_1+\bar{o}_1-\bar{g}_1 \ ,\nonumber \\
D_{g;f} &=& i(o_1+g_1-\bar{o}_1-\bar{g}_1) \ , \hspace{1cm} 
D_{g;h}  \ \ = \ \  i(o_1-g_1-\bar{o}_1+\bar{g}_1) \ ,
\nonumber \\
D_{f;o} &=& o_2+g_2+\bar{o}_2+\bar{g}_2 \ , \hspace{1.5cm} 
D_{f;g}  \ \ =  \ \ i(o_2+g_2-\bar{o}_2-\bar{g}_2) \ ,\nonumber \\
D_{f;f} &=& o_2-g_2+\bar{o}_2-\bar{g}_2 \ , \hspace{1.5cm} 
D_{f;h}  \ \ =  \ \ -i(o_2-g_2-\bar{o}_2+\bar{g}_2) \ ,
\nonumber \\
D_{h;o} &=& o_3+g_3+\bar{o}_3+\bar{g}_3 \ , \hspace{1.5cm} D_{h;g} 
 \ \ =  \ \ i(o_3+g_3-\bar{o}_3-\bar{g}_3) \ ,
\nonumber \\
D_{h;f} &=& i(o_3-g_3-\bar{o}_3+\bar{g}_3) \ , \hspace{1cm} 
D_{h;h}  \ \ =  \ \ o_3-g_3+\bar{o}_3-\bar{g}_3 \ .
\label{a22}
\ea
The D9 branes have ${\cal N}=1$ supersymmetry, with gauge 
group ${\rm SO}(8)^4$ and chiral multiplets in 
the $(8,8,1,1)$ and five permutations.
Moreover, each antibrane gives a non-supersymmetric 
spectrum, with gauge group ${\rm U}(8) \times {\rm U}(8)$, chiral 
multiplets in the $(8,8)$, 
$(8,\bar{8})$ and in their conjugates, spinors in 
the $(28,1)$, $(\overline{28},1)$, 
$(1,28)$, $(1,\overline{28})$ and complex scalars in 
the $(36,1)$, $(\overline{36},1)$, $(1,36)$, $(1,\overline{36})$.
We would like to stress that in this case the gauginos are massless,
since the M{\"o}bius amplitude does not affect the adjoint 
representations of unitary 
groups. Finally, $\bar{5}_k\bar{5}_l$ sectors give chiral multiplets
in the representations
\ba
\bar{5}_1\bar{5}_2  & & (8,1;8,1) \ , \ \ (\bar{8},1;1,8) \ , 
          \ \ (1,8;\bar{8},1) \ , \ \ (1,\bar{8};1,\bar{8}) \ , 
\nonumber \\
\bar{5}_1\bar{5}_3  & & (8,1;\bar{8},1) \ , \ \ (\bar{8},1;1,\bar{8}) \ , 
          \ \ (1,8;1,8) \ , \ \ (1,\bar{8};8,1) \ , 
\nonumber \\
\bar{5}_2\bar{5}_3   & &   (8,1;\bar{8},1) \ , \ \ (\bar{8},1;1,8) \ , 
          \ \ (1,8;1,\bar{8}) \ , \ \ (1,\bar{8};8,1) \ ,
\nonumber
\ea 
and the non-supersymmetric $9\bar{5}_k$ sectors contain Weyl spinors
and complex scalars in the representations
\ba
9\bar{5}_1 &{\it \ {\rm spinors} \ :}& \
  (8,1,1,1;8,1) \ , \ \ (1,8,1,1;\bar{8},1) \ , 
          \ \ (1,1,8,1;1,8) \ , \ \ (1,1,1,8;1,\bar{8}) \ , 
\nonumber \\
           &{\it \ {\rm scalars} \ :}&  \ 
(8,1,1,1;1,8) \ , \ \ (1,8,1,1;1,\bar{8}) \ , 
          \ \ (1,1,8,1;8,1) \ , \ \ (1,1,1,8;\bar{8},1) \ , 
\nonumber \\
9\bar{5}_2 &{\it \ {\rm spinors} \ :}& \
  (8,1,1,1;8,1) \ , \ \ (1,8,1,1;1,8) \ , 
          \ \ (1,1,8,1;\bar{8},1) \ , \ \ (1,1,1,8;1,\bar{8}) \ , 
\nonumber \\
           &{\it \ {\rm scalars} \ :}&  \ 
(8,1,1,1;1,8) \ , \ \ (1,8,1,1;8,1) \ , 
          \ \ (1,1,8,1;1,\bar{8}) \ , \ \ (1,1,1,8;\bar{8},1) \ , 
\nonumber \\
9\bar{5}_3 &{\it \ {\rm spinors} \ :}& \
  (8,1,1,1;\bar{8},1) \ , \ \ (1,8,1,1;1,\bar{8}) \ , 
          \ \ (1,1,8,1;1,8) \ , \ \ (1,1,1,8;8,1) \ , 
\nonumber \\
           &{\it \ {\rm scalars} \ :}&  \ 
(8,1,1,1;1,\bar{8}) \ , \ \ (1,8,1,1;\bar{8},1) \ , 
          \ \ (1,1,8,1;8,1) \ , \ \ (1,1,1,8;1,8) \ . \nonumber
\ea
All chiral spectra thus obtained are free of non-abelian anomalies. 

On the other hand, the models without discrete torsion are not chiral. 
The choice $(\e_1,\e_2,\e_3) = (+++)$,
discussed in \cite{erice} and worked out in detail in \cite{bl}, 
leads to a gauge group ${\rm USp}(16)^4$. Another model, without 
discrete torsion 
but with two D${\bar 5}$ branes, can be obtained
letting two of the $\e_k$ be negative. The D9 and D5 branes give 
orthogonal gauge groups with ${\cal N}=1$ supersymmetry, while the two 
D${\bar 5}$ branes give symplectic gauge groups with no supersymmetry. For 
instance, with the choice $(+--)$
\ba
{\cal A}_0 &=& \frac{1}{8} \biggl \{ 
(N_o^2+ D^2_{g;o}+ D^2_{f;o}+ D^2_{h;o})T_{oo} 
+2N_oD_{g;o}T_{go}+2N_oD_{f;o}T^{(-)}_{fo}+2N_oD_{h;o}T^{(-)}_{ho} \nonumber \\
&+& 2D_{g;o}D_{f;o}T^{(-)}_{ho}+2D_{g;o}D_{h;o}T^{(-)}_{fo}
+2D_{f;o}D_{h;o}T_{go} \biggr \}  \quad , \nonumber \\
{\cal M}_0 &=& - \frac{1}{4} \biggl\{ 
(N_o +D_{g;o}) (\t_{oo}-\t_{og}+\t_{of}+\t_{oh})
\nonumber 
\\
&-& (D_{f;o}+D_{h;o}) \left[ 
(\t^{{\rm NS}}_{oo}-\t^{{\rm NS}}_{og}+\t^{{\rm NS}}_{of}+\t^{{\rm NS}}_{oh})
+(\t^{{\rm R}}_{oo}-\t^{{\rm R}}_{og}+\t^{{\rm R}}_{of}+
\t^{{\rm R}}_{oh})\right] 
\biggr \} \quad  ,
\label{a26}
\ea
and there are no breaking terms in the annulus. After a suitable rescaling
of the charge multiplicities,
the D$\bar{5}_2$ and the D$\bar{5}_3$ branes give non-supersymmetric spectra, 
with ${\rm USp}(16)$ gauge groups, spinors in the 136 and 
in three copies of the 120
and scalars in the 120 and in two copies of the 136. 
The 99 and $5_15_1$ sectors
have ${\cal N}=1$ supersymmetry, gauge group ${\rm SO} (16)$ and 
chiral multiplets in the
136 and in two copies of the 120.
Finally, there are two chiral multiplets in the representation
$(16,16)$ arising from the $95_1$ and the $\bar{5}_2 \bar{5}_3$
sectors and complex scalars and Weyl spinors in bifundamental
representations arising from the $9\bar{5}_2$, $9\bar{5}_3$, 
$5_1 \bar{5}_2$ and $5_1 \bar{5}_3$ sectors. 

The fact that the $Z_2 \times Z_2$ orientifold with discrete torsion naturally
leads to non-supersymmetric spectra can also be argued considering
F-theory on the $T^8/Z_2 \times Z_2 \times Z_2$
with discrete torsion \cite{kakul}, since the blow-up of this 
eight-dimensional orbifold is only birational to a Calabi-Yau fourfold
\footnote{We would like to thank 
Zurab Kakushadze and Koushik Ray for calling this correspondence to
our attention.}.

\subsection{Comments on freely acting $Z_2 \times Z_2$ orientifolds with brane 
supersymmetry breaking}

In all the models discussed in \cite{adds2}, one can introduce 
D${\bar 5}$ branes
simply reverting some signs in the twisted sectors of the Klein bottle, as in 
the previous case. This procedure generates models with branes, antibranes
and various amounts of supersymmetry. For instance, the $p_{23}$ model
 of \cite{adds2}, with $\frac{1}{2}$-momentum shifts along the last two tori,
can be turned into a different model, described by
\ba
{\cal K} &=& \frac{1}{8} \biggl\{ T_{oo} \bigl[ P_1P_2P_3 + P_1W_2W_3 + 
W_1(-1)^{m_2}P_2W_3 + W_1W_2(-1)^{m_3}P_3 \bigr] \nonumber \\
&-& 2 \times 16 T_{go}\ P_1 
\left(\frac{ \eta}{\vartheta_4}\right)^2 \biggr\}  \quad , 
\nonumber \\
{\cal A} &=& \frac{1}{8} \Biggl\{ 
T_{oo} \left[ N_o^2P_1P_2P_3 + \frac{D_{g;o}^2}{2}P_1  \left( W_2W_3 \!+\! 
W_2^{n+1/2}W_3^{n+1/2}\right) \right] 
\nonumber \\
&+& (N_g^2 + 2 D_{g;g}^2) T_{og}P_1
\left( \frac{2 \eta}{\vartheta_2}\right)^2  
+
N_f^2 \ T_{of}(-1)^{m_2}P_2\left( \frac{2 \eta}{\vartheta_2}
\right)^2 
+ N_h^2 \ T_{oh}(-1)^{m_3}P_3\left( \frac{2 \eta}{\vartheta_2}\right)^2
\nonumber \\   
&+& 2 N_o D_{g;o} T^{(-)}_{go}P_1\left( \frac{ \eta}{\vartheta_4}\right)^2  
- 4N_gD_{g;g}T^{(-)}_{gg}P_1\left( \frac{ \eta}{\vartheta_3}\right)^2  
\Biggr\} \quad , 
\nonumber \\
{\cal M} &=& - \frac{1}{8} \Biggl\{ \left( \hat{T}_{oo}N_oP_1P_2P_3 -  
\hat{\tilde T}{}_{oo}^{(-)}D_{g;o} P_1W_2W_3 \right)
+ \left( \hat{T}_{og}N_o - \hat{\tilde T}{}_{og}^{(-)}D_{g;o} \right)
P_1\left( \frac{2 \hat{\eta}}{\hat{\vartheta_2}}\right)^2 \nonumber 
\\ 
&+& N_o\hat{T}_{of}(-1)^{m_2}P_2
\left( \frac{2 \hat{\eta}}{\hat{\vartheta_2}}\right)^2    
+ N_o\hat{T}_{oh}(-1)^{m_3}
P_3\left( \frac{2 \hat{\eta}}{\hat{\vartheta_2}}\right)^2  
 \biggr\} \quad .
\label{f1}
\ea
The 99 sector gives a gauge group ${\rm SO}(16-a)\times 
{\rm SO}(a) \times {\rm  SO}(16-c) 
\times {\rm SO}(c)$, with ${\cal N}=1$ supersymmetry and chiral 
multiplets in the 
representation $(16-a,a,1,1)$ and in five additional ones 
differing by permutations of the entries.
The $\bar{5}_1\bar{5}_1$ sector gives a gauge group ${\rm USp}(8) \times 
{\rm USp}(8)$, with 
complex scalars in the $(36,1)$ and $(1,36)$, Weyl spinors in two
copies of the $(28,1)$ and 
$(1,28)$, and one ${\cal N}=2$ hypermultiplet in the $(8,8)$.
Finally, the $9\bar{5}_1$ sector gives Weyl spinors in the representations
$(16-a,1,1,1;8,1)$, $(1,a,1,1;8,1)$, $(1,1,16-c,1;1,8)$, $(1,1,1,c;1,8)$
and complex scalars in the representations
$(16-a,1,1,1;1,8)$, $(1,a,1,1;1,8)$, $(1,1,16-c,1;8,1)$, $(1,1,1,c;8,1)$.
In the decompactification limit $R_2,R_3 \rightarrow \infty$, local tadpole 
cancellation requires a further breaking of the D${\bar 5}$ gauge group to 
${\rm USp}(4)^4$, and the resulting configuration may be linked to the $Z_2$ 
orientifold discussed in \cite{ads2}.

\section{Open descendants of the $T^6 / Z_4$ orbifold} 

The $Z_4$ orbifold is obtained identifying the 
complex world-sheet coordinates $X^a$ on the three internal tori and the
world-sheet fermions $\lambda^a$ according to 
$X^a \sim \w_a X^a $, $\lambda^a \sim \w_a \lambda^a $, 
where $\w_a = e^{2 \pi i t_a}$  $(a=1,2,3)$ and the 
twist vector has components  $(t_1,t_2,t_3)=(\frac{1}{4},\frac{1}{4},-
\frac{1}{2})$. In order to display the 
contributions of the $\lambda^a$, it is convenient to 
decompose the original level-one SO(8) characters 
with respect to  
${\rm SO}(2) \times {\rm SO}(2) \times {\rm SU}(2) \times {\rm U}(1)$. 
To this end, we introduce the two level-one SU(2) 
characters $(\chi_0,\chi_{1/2})$, of conformal weights
$(0,\frac{1}{4})$, and the 
eight U(1) characters $\x_m$  $(m=0,\pm 1,\pm 2,\pm 3, 4)$ for a boson 
on a circle of radius $R= \sqrt{8}$, of conformal weights
$h_m = m^2/16$. The  contribution of the $\lambda^a$
to the sector twisted by $\w^k$ and projected by $\w^l$, $(k,l=0,1,2,3)$, 
can then be expressed as
\be
\psi_{kl} = \rho_{k0}+ i^l \rho_{k1}+ (-1)^l \rho_{k2}+ (-i)^l \rho_{k3} 
\quad ,
\label{q1}
\ee
where the characters $\rho_{kl}$ are collected in Appendix B,
while the 
contributions of the internal (lattice) bosons are
\ba
\f_{kl} = \left[-2 \sin \left(\frac{\pi l}{4}
\right)\right]^{\delta_{k,0}} \,
\frac{\h}{\theta \left[ {1/2 + k/4} \atop {1/2 + l/4} \right] } \ .
\label{q2}
\ea
The torus amplitude is then
\ba
{\cal T} &=& \frac{1}{4} \biggl \{ |\psi_{00}|^2\L_1\L_2\L_3 
+ \psi_{01}\bar{\psi}_{03}\f_{01}^2 \bar{\f}^2_{03}
\left| \frac{2\eta}{\vartheta_2} \right|^2 + 
|\psi_{02}|^2\left| \frac{2\eta}{\vartheta_2} \right|^4\L_3
+\psi_{03}\bar{\psi}_{01}\f_{03}^2 \bar{\f}^2_{01}
\left| \frac{2\eta}{\vartheta_2} \right|^2
\nonumber \\
&+& 16 \biggl ( |\psi_{20}|^2\left| \frac{\eta}{\vartheta_4} \right|^4\L_3
+|\psi_{22}|^2 \left| \frac{\eta}{\vartheta_3} \right|^4\L_3 \biggr )
+ 4 \biggl ( \psi_{21}\bar{\psi}_{23}\f_{21}^2\bar{\f}_{23}^2\left| 
\frac{2\eta}{\vartheta_2} \right|^2
+\psi_{23}\bar{\psi}_{21}\f_{23}^2\bar{\f}_{21}^2\left| 
\frac{2\eta}{\vartheta_2} \right|^2 \biggr ) \nonumber \\
&+& 16 \biggl(\psi_{10}\bar{\psi}_{30}\f_{10}^2\bar{\f}_{30}^2\left| 
\frac{\eta}{\vartheta_4} \right|^2
+\psi_{11}\bar{\psi}_{33}\f_{11}^2\bar{\f}_{33}^2\left| 
\frac{\eta}{\vartheta_3} \right|^2
+\psi_{12}\bar{\psi}_{32}\f_{12}^2\bar{\f}_{32}^2\left| 
\frac{\eta}{\vartheta_4} \right|^2
\nonumber \\
&+& \psi_{13}\bar{\psi}_{31}\f_{13}^2 \bar{\f}_{31}^2  \left| 
\frac{\eta}{\vartheta_3} \right|^2
+ \psi_{30}\bar{\psi}_{10}\f_{30}^2\bar{\f}_{10}^2\left| 
\frac{\eta}{\vartheta_4} \right|^2
+\psi_{31}\bar{\psi}_{13}\f_{31}^2\bar{\f}_{13}^2\left| 
\frac{\eta}{\vartheta_3} \right|^2
\nonumber \\
&+& \psi_{32}\bar{\psi}_{12}\f_{32}^2\bar{\f}_{12}^2\left| 
\frac{\eta}{\vartheta_4} \right|^2
+\psi_{33}\bar{\psi}_{11}\f_{33}^2 \bar{\f}_{11}^2  \left| 
\frac{\eta}{\vartheta_3} \right|^2 
\biggr ) \biggl \} \quad .
\label{q3}
\ea
Before the $\W$ projection, the massless sector describes ${\cal N}=2$ 
supergravity 
coupled to $7$ vector multiplets and $32$ hypermultiplets, and can thus be 
associated with the singular limit of a Calabi-Yau manifold with 
Hodge numbers $(h_{11},h_{21})=(31,7)$.

Let us now turn to the construction of the open descendants, starting 
from the Klein bottle amplitude
\ba
{\cal K} &=& \frac{1}{8} \biggl \{ \psi_{00}\biggl(P_1P_2P_3+W_1W_2P_3\biggr)
+ 2 \, \psi_{02}\left( \frac{2\eta}{\vartheta_2} \right)^2W_3  \nonumber \\
&+& \e \biggl [ 2 \times 16 \, \psi_{20}\left( \frac{\eta}{\vartheta_4} 
\right)^2P_3
+ 2 \times 4 \, \psi_{22}\left( \frac{\eta}{\vartheta_3} \right)^2W_3 
\biggr ]\biggr \} \ ,
\label{q4}
\ea
where, as in the previous cases, we have inserted a sign $\e = \pm 1$ 
in front of the twisted contributions. The choice 
$\e=1$ gives the usual $\W$ projection, while the choice $\e=-1$ 
inverts the charge of the O5 plane and, as in the previous examples, requires 
the introduction of antibranes. In the transverse channel, this amplitude 
turns into
\ba
\tilde{{\cal K}} &=& \frac{2^5}{8} \biggl \{ 
\psi_{00}\biggl(v_1v_2v_3W^e_1W^e_2W^e_3+\frac{v_3}{v_1v_2}
P^e_1P^e_2W^e_3 \biggr)
+2 \, \psi_{20}\left( \frac{\eta}{\vartheta_4} \right)^2\frac{P^e_3}{v_3}
\nonumber \\
&+& 2 \, \e \, \psi_{02}\left( \frac{2\eta}{\vartheta_2} \right)^2{W^e_3}v_3
-2 \, \e \, \psi_{22}\left( \frac{\eta}{\vartheta_3} \right)^2\frac{P^e_3}{v_3}
\biggr \} \quad ,
\label{q5}
\ea
and a closer look at the contributions at the origin of the internal lattices
\ba
\tilde{{\cal K}}_0 &=& \frac{2^5}{8} \biggl \{
\biggl (\sqrt{v_1v_2v_3}+\e \sqrt{\frac{v_3}{v_1v_2}} \biggr )^2
(\rho_{00}+\rho_{02})
+\biggl (\sqrt{v_1v_2v_3}-\e \sqrt{\frac{v_3}{v_1v_2}} \biggr )^2
(\rho_{01}+\rho_{03}) \nonumber \\
&+& \frac{4}{v_3}\frac{1}{2} \left[ (1-\e) (\rho_{20}+\rho_{22})
+ (1+\e) (\rho_{21}+\rho_{23}) \right] \biggr \}
\label{q6}
\ea
reveals the presence of a term proportional to $1/v_3$. For $\e=1$,
this gives rise to a massless tadpole that can not be canceled
by the annulus and M{\"o}bius contributions. 
For $\e=-1$, however, this term becomes
massive, and one can complete the construction of the open descendants
without any further difficulties.
In this case, the projected massless closed spectrum
is ${\cal N}=1$ supergravity coupled to $6$ vector
multiplets and $33$ chiral multiplets.  For the open sector,
it is convenient to introduce a compact notation, as in the previous Section, 
defining
\be
\tilde{\psi}_{kl}^{(\varepsilon)}=
\psi^{\rm NS}_{kl}- \varepsilon \, \psi^{\rm R}_{kl} \quad ,
\label{q7}
\ee
and denoting by 
\be
\psi^{(-)}_{kl} = \s_{k0}+ i^l\s_{k1}+ (-1)^l \s_{k2}+ (-i)^l \s_{k3}  
\label{q7b}
\ee
the combinations of characters obtained from $\tilde{\psi}_{lk}^{(-)}$ 
after an $S$ modular transformation. The explicit definition of the 
$\s_{kl}$ may be found in Appendix B.

The transverse-channel annulus amplitude is then
\ba
\tilde{{\cal A}} &=& \frac{2^{-5}}{8}\biggl \{ 
(N^2v_1v_2v_3W_1W_2W_3+D^2\frac{v_3}{v_1v_2}P_1P_2W_3)\psi_{00}
+32 (R^2+R_D^2)\psi_{30}\f_{30}^2\left( \frac{\eta}{\vartheta_4} \right) 
\nonumber \\
&+& 16(S^2+S_D^2)\psi_{20}\left( \frac{\eta}{\vartheta_4} \right)^2v_3W_3 
+ 32 (T^2+T_D^2)\psi_{10}\f_{10}^2\left( \frac{\eta}{\vartheta_4} \right) 
\nonumber \\
&+& 2ND{\tilde \psi}_{02}^{(-)}\left( \frac{2\eta}{\vartheta_2} \right)^2v_3W_3
+32RR_D\tilde \psi_{32}^{(-)}\f_{32}^2\left( \frac{\eta}{\vartheta_4} \right) 
\nonumber \\
&-& 8SS_D\tilde \psi_{22}^{(-)}\left( \frac{\eta}{\vartheta_3} \right)^2v_3W_3
+32TT_D\tilde \psi_{12}^{(-)}\f_{12}^2\left( \frac{\eta}{\vartheta_4} \right) 
\biggr \} \label{q9}
\ea
and, in particular, the untwisted contributions at the origin 
of the lattices are
\ba
\tilde{{\cal A}}_0 &=& \frac{2^{-5}}{8} \biggl \{ 
\left( N\sqrt{v_1v_2v_3}+D\sqrt{\frac{v_3}{v_1v_2}} \right)^2
(\rho_{00}+\rho_{02})^{\rm NS} 
\nonumber \\
&+& \left( N\sqrt{v_1v_2v_3}-D\sqrt{\frac{v_3}{v_1v_2}} \right)^2
(\rho_{01}+\rho_{03})^{\rm NS} 
\nonumber \\
&-&\left(N\sqrt{v_1v_2v_3}-D\sqrt{\frac{v_3}{v_1v_2}} \right)^2
(\rho_{00}+\rho_{02})^{\rm R} 
\nonumber \\
&-& \left(N\sqrt{v_1v_2v_3}+D\sqrt{\frac{v_3}{v_1v_2}} \right)^2
(\rho_{01}+\rho_{03})^{\rm R} \biggr \} \quad .
\label{q10}
\ea
It should be appreciated that the structure of the breaking terms is 
consistent with eq. $(\ref{a15b})$. Thus, in the sector twisted by $\w^2$
they are
\be
(S+4S_D)^2+15S^2 = 16S^2+16S_D^2+8SS_D \quad ,
\label{q11}
\ee
since the branes are present only in one fixed torus, while in the sector 
twisted by $\w$ (or $\w^3$) they are
\be
4(R+2R_D)^2+12R^2 = 16R^2+16R_D^2+16RR_D \quad,
\label{q12}
\ee
since the branes fill the third torus, and are thus present in 
four of the $\w$-fixed points.

As usual, the transverse-channel amplitudes $\tilde{\cal K}$ and 
$\tilde{\cal A}$ determine
\ba
\tilde{{\cal M}} &=& - \frac{1}{4} \Biggl \{ Nv_1v_2v_3W^e_1W_2^eW_3^e
\hat{\psi}_{00}-Nv_3W^e_3\hat{\psi}_{02}
\left( \frac{2\hat{\eta}}{\hat{\vartheta_2}}\right )^2
-D\frac{v_3}{v_1v_2}P^e_1P^e_2W^e_3\hat{\tilde \psi}{}_{00}^{(-)} 
\nonumber \\
&+& Dv_3W^e_3\hat{\tilde \psi}{}_{02}^{(-)}
\left( \frac{2\hat{\eta}}{\hat{\vartheta_2}} \right )^2
+ 2\left( S\hat{\psi}_{21}-S_D\hat{\tilde \psi}{}_{21}^{(-)}\right)
\hat{\f}_{21}^2
\left( \frac{2\hat{\eta}}{\hat{\vartheta_2}} \right) 
\nonumber \\
&+& 2\left( S\hat{\psi}_{23}-S_D\hat{\tilde \psi}{}_{23}^{(-)}\right )
\hat{\f}_{23}^2
\left( \frac{2\hat{\eta}}{\hat{\vartheta_2}} \right) 
\Biggr \} \quad , \label{qm10}
\ea
and the tadpole conditions, now solvable, are
\be
N = D = 32 \ , \hspace{1cm} R=R_D=S=S_D=T=T_D=0 \ .
\label{q15}
\ee
After $S$ and $P$ transformations, one can then recover from 
(\ref{q9}) and (\ref{qm10}) the direct-channel 
amplitudes
\ba
{\cal A} &=& \frac{1}{8} \Biggl\{ 
\left( N^2P_1P_2P_3+D^2W_1W_2P_3 \right) \psi_{00}
+\left( R^2+R_D^2 \right) \psi_{01}
\f_{01}^2\left( \frac{2\eta}{\vartheta_2} \right)
\nonumber \\
&+& \left(S^2+S_D^2\right)\psi_{02}
\left( \frac{2\eta}{\vartheta_2} \right)^2P_3
+ \left(T^2+T_D^2\right)\psi_{03}\f_{03}^2\left(
\frac{2\eta}{\vartheta_2} \right) + 2ND\psi^{(-)}_{20}\left(
\frac{\eta}{\vartheta_4} \right)^2P_3 
\nonumber \\
&+& 2RR_D \psi^{(-)}_{21}\f_{21}^2\left(
\frac{2\eta}{\vartheta_2} \right) +
2SS_D\psi^{(-)}_{22}\left( \frac{\eta}{\vartheta_3} \right)^2P_3
+2TT_D\psi^{(-)}_{23}\f_{23}^2\left( \frac{2\eta}{\vartheta_2} \right)
\biggr \} 
\label{q8}
\ea
and
\ba
{\cal M} &=& - \frac{1}{8} \biggl \{ N P_1 P_2 P_3
\hat{\psi}_{00} + N P_3 \hat{\psi}_{02} \left( 
\frac{2\hat{\eta}}{\hat{\vartheta_2}}
\right )^2 - D W_1 W_2 P_3 \hat{\tilde\psi}{}_{00}^{(-)}
- D P_3 \hat{\tilde\psi}{}_{02}^{(-)}
\left( \frac{2\hat{\eta}}{\hat{\vartheta_2}} \right )^2 
\nonumber 
\\
\!\!\!\!&+&\!\!\!\! S \hat{\psi}_{01} \hat{\f}_{01}^2
\left( \frac{2\hat{\eta}}{\hat{\vartheta_2}} \right) 
\!+\! S \hat{\psi}_{03} \hat{\f}_{03}^2
\left( \frac{2\hat{\eta}}{\hat{\vartheta_2}} \right) 
\!-\! S_D \hat{\tilde\psi}{}_{01}^{(-)} \hat{\f}_{01}^2
\left( \frac{2\hat{\eta}}{\hat{\vartheta_2}} \right) 
\!-\! S_D \hat{\tilde\psi}{}_{03}^{(-)} \hat{\f}_{03}^2
\left( \frac{2\hat{\eta}}{\hat{\vartheta_2}} \right) 
\biggr \}  \, .
\label{q13}
\ea
As usual, the
coefficients in $\tilde{\cal M}$ are fixed by factorization in
the tube channel. Some sign ambiguities apparently present for
massive characters, whose coefficients are not constrained by tadpoles,
are fixed demanding that the direct-channel amplitudes
have a correct particle interpretation. The latter calls for the
parametrization
\ba
&N =  n+m+p+\bar{m} \ , \qquad &\,\,\, D = d+r+q+\bar{r} \ ,
\nonumber \\
&R =  n+im-p-i\bar{m} \ ,  \qquad &R_D = d+ir-q-i\bar{r} \ , 
\nonumber \\ 
&S = n-m+p-\bar{m} \ ,  \qquad &S_D = d-r+q-\bar{r} \ ,
\nonumber \\
&T = n-im-p+i\bar{m} \ ,  \qquad &T_D = d-ir-q+i\bar{r} \ ,
\label{q14}
\ea
that implements the $Z_4$ orbifold projections on the Chan-Paton charges.

One can now read the massless open spectrum from the amplitudes 
$(\ref{q8})$, $(\ref{q13})$ restricted to the origin of 
the internal lattice
\ba
{\cal A}_0+ {\cal M}_0 &=& \frac{1}{2} \biggl\{ 
\rho_{00} \left(n^2+p^2+2m\bar{m}+d^2+q^2+ 2r\bar{r} \right)
+\rho_{01} \left(2n\bar{m}+2pm+2d\bar{r}+2qr\right) 
\nonumber \\
&+& \rho_{02}\left(2np+m^2+\bar{m}^2+2dq+r^2+\bar{r}^2 \right)
+\rho_{03} \left(2nm+2p\bar{m}+2dr+2q\bar{r}\right) 
\nonumber \\
&+&\s_{20}\left(nd+pq+m\bar{r}+\bar{m}r\right)
+\s_{21}\left(n\bar{r}+mq+pr+\bar{m}d \right)
\nonumber \\
&+& \s_{22}\left(nq+mr+pd+\bar{m}\bar{r}\right)
+ \s_{23}\left(nr+md+p\bar{r}+\bar{m}q \right) \biggr\} \nonumber \\
&-& \frac{1}{2} 
\left\{ \hat{\rho}_{00}(n+p)-\hat{\tilde\rho}{}_{00}^{(-)}(d+q)
+\hat{\rho}_{02}(m+\bar{m})-\hat{\tilde\rho}{}_{02}^{(-)}
(r+\bar{r}) \right\} \quad ,
\ea
while the tadpole conditions select the gauge group 
${\rm SO} (8)_9$ $\times$ ${\rm SO}(8)_9$ $\times$ ${\rm U}(8)_9$ 
$\times$ ${\rm USp}(8)_{\bar 5}$ $\times$ ${\rm USp}(8)_{\bar 5}$ $\times$
$ {\rm U}(8)_{\bar 5}$. 
The D9 spectrum has ${\cal N}=1$ 
supersymmetry, with two chiral multiplets in the $(8,1,\bar{8})$
and in the $(1,8,8)$ representations, and one chiral multiplet in each of the 
$(1,1,28)$, $(1,1,\overline{28})$ and $(8,8,1)$ representations.
On the other hand, the D$\bar{5}$ spectrum is not supersymmetric, 
and contains, aside from the corresponding gauge bosons, spinors 
in the $(28,1,1)$, $(1,28,1)$, $(1,1,64)$, 
$(1,1,28)$, $(1,1,\overline{28})$, complex
scalars in the $(1,1,36)$, $(1,1,\overline{36})$, and chiral 
multiplets in the $(8,8,1)$ and in two copies of the
$(8,1,\bar{8})$ and $(1,8,8)$.
Finally, the $ND$ strings contain complex scalars in the
$(8,1,1;8,1,1)$, $(1,8,1;1,8,1)$, $(1,1,8;1,1,\bar{8})$ and 
$(1,1,\bar{8};1,1,8)$, and Weyl spinors in the
$(8,1,1;1,1,\bar{8})$, $(1,1,8;1,8,1)$, $(1,8,1;1,1,8)$ 
and $(1,1,\bar{8};8,1,1)$.
This spectrum is chiral, but is free of irreducible gauge anomalies.
As in previous examples, if suitable diagonal subgroups of the 
D${\bar 5}$ factors are regarded as global symmetries, the D9 
spectrum  has ${\cal N}=1$ supersymmetry.


\section{Type I vacua with branes and antibranes of the same type}

In the previous examples we have seen how the structure of
the closed sector determines both the types and the 
total numbers of D-branes present in the open descendants. 
In these cases, the breaking of 
supersymmetry on the branes is directly enforced by the consistency
of the model. 

Somewhat different scenarios have been recently proposed in \cite{su,au,aiq}.
In the resulting models, a supersymmetric open sector is deformed 
allowing for the simultaneous 
presence of branes and antibranes of the same type. Whereas tadpole
conditions only fix the total RR charge, the option of saturating it by 
a single type of D-brane, whenever available, stands out as the only one 
compatible with space-time supersymmetry.
However, if one relaxes this last condition, there are no 
evident obstructions to
considering vacuum configurations where branes and antibranes with
a fixed total RR charge are simultaneously present. The rules for 
constructing this wider class of models can be simply presented referring 
to a ten-dimensional example. 

The starting point is the familiar supersymmetric type IIB torus amplitude 
\begin{equation}
{\cal T} = |V_8 - S_8 |^2 \quad ,
\end{equation}
and the corresponding Klein bottle projection
\begin{equation}
{\cal K} = \frac{1}{2} (V_8 - S_8) \quad .
\end{equation}
In the transverse channel, the latter becomes 
\begin{equation}
\tilde{\cal K} = \frac{2^5}{2} (V_8 - S_8) \quad ,
\end{equation} 
and requires an open sector with a {\it net} number
of 32 branes in order to cancel the resulting RR tadpole.
Actually, both $V_8$ and $S_8$ develop tadpoles in this case,
that in the usual type I model are related by supersymmetry, 
but are conceptually quite different. While NS-NS tadpoles result in
redefinitions of the vacuum configuration, RR tadpoles signal in general 
genuine inconsistencies, and their presence is a symptom of the
emergence of serious pathologies \cite{pc}.

Actually, one can conceive a more general construction in this case \cite{su}, 
allowing in the transverse-channel annulus different reflection 
coefficients for
the $V_8$ and $S_8$ characters, so that
\begin{equation}
\tilde{\cal A} = \frac{2^{-5}}{2} \left[ (n_+ + n_-) ^2 \, V_8 - (n_+ - n_-
)^2 \, S_8 \right] \ ,
\end{equation} 
where $n_+$ and $n_-$ actually count the total numbers of D9 and 
D$\bar 9$ branes. It should be appreciated how their relative minus sign in 
the coefficient 
of $S_8$ accounts neatly for their opposite RR charges, while they 
have clearly identical couplings to the graviton, consistently with
the coefficient of $V_8$. The direct-channel annulus amplitude
\begin{equation}
{\cal A} = \frac{n_{+}^{2} + n_{-}^{2}}{2}\, (V_8 - S_8) + n_+ n_- \,
(O_8 - C_8)
\end{equation}
reflects the opposite GSO projections for open
strings stretched between two D-branes of the same type (99 or
$\bar 9\bar 9$) and of different types ($9\bar 9
$ or $\bar 9 9$) \cite{sen}. While the former yields
the supersymmetric type I spectrum, the latter eliminates the vector and 
its spinorial superpartners, while retaining the
tachyon and the spinor of opposite chirality. As a
result, supersymmetry is broken and an instability, signaled
by the presence of the tachyonic ground state, emerges. 

The M{\"o}bius amplitude 
\begin{equation}
{\cal M} = \pm \frac{1}{2} (n_+ + n_-) \, \hat V _8 + \frac{1}{2} (n_+ -
n_-) \, \hat S _8
\end{equation}
now involves naturally an undetermined sign for $V_8$, whose
tadpole is generally incompatible with the one of $S_8$, and
is to be relaxed. Together with ${\cal A}$, the two signs lead to
symplectic or orthogonal gauge groups with $S$ fermions in (anti)symmetric 
representations and tachyons and $C$ fermions in bi-fundamentals.

In these ten-dimensional models, the only way to eliminate the tachyon
consists in introducing only D9-branes. Depending on
the signs in the M{\"o}bius amplitude, one thus recovers either
the SO(32) superstring or the USp(32) model of \cite{su}. 
On the other hand, more can be done if one
compactifies the theory on some internal manifold. In this case, one
can introduce Wilson lines (or, equivalently, separate the branes) 
in such a way that in the open strings 
stretched between separate 9 and $\bar 9$ branes the tachyon 
actually becomes massive. 
It is instructive to analyze in some detail the simple case of circle 
compactification.
As in the previous example, the Klein bottle amplitude is not affected, 
and is given by
\begin{equation}
{\cal K} = \frac{1}{2} \, (V_8 - S_8 ) \, P \ ,
\end{equation}
where $P$ denotes the sum over momentum states.
However, the Wilson line affects the annulus amplitude, that in the
transverse-channel now reads
\begin{equation}
\tilde{\cal A}  = \frac{2^{-5}}{2} \left[ (n_+ + (-1)^n n_- )^2 \, V_8 -
(n_+ - (-1)^n n_- )^2 \, S_8 \right]\, W \quad ,
\end{equation}
where $(-1)^n W$ denotes an oscillating winding sum.
As a result, in the direct channel amplitude
\begin{equation}
{\cal A} = \frac{n_{+}^{2} + n_{-}^{2}}{2} \, (V_8 - S_8)\, P +
n_+ n_- \,(O_8 - C_8)\, P^{(1/2)} \ ,
\end{equation}
where $P^{(1/2)}$
denotes a sum over $\frac{1}{2}$-shifted momentum states,
both the tachyon and the $C$ spinor are lifted.
The open sector is completed by the M{\"o}bius amplitude 
\begin{equation}
{\cal M} = \frac{1}{2} \left[ \pm (n_+ + n_- )\, \hat V_8 + (n_+ -
n_-) \, \hat S_8 \right]  P \ ,
\end{equation}
and at the massless level comprises gauge bosons in the adjoint of
${\rm SO}(n_+) \times {\rm SO}(n_-)$ (or  
${\rm USp}(n_+) \times {\rm USp}(n_-)$, depending on the sign of
$\hat V _8$ in ${\cal M}$) and $S$ spinors in 
(anti)symmetric representations.

\subsection{Toroidal compactifications with nine and five (anti)branes}

Let us now turn to six-dimensional toroidal compactifications. In this
case, one has the interesting option to introduce in 
the standard type I model of
\cite{toroidal} pairs of D5--D$\bar 5$ and D9--D$\bar 9$
branes. The result includes a {\it chiral} spectrum confined to the
(non-supersymmetric) branes, and calls for the introduction of
six-dimensional Green-Schwarz couplings to the single tensor 
present in the projected closed spectrum to cancel the residual gauge 
and mixed anomalies.

The starting point in this construction is the standard type IIB string
compactified on a four-torus, for which
\begin{equation}
{\cal T} = |V_8 - S_8 |^2 \, \Lambda^4
= |V_4 O_4 + O_4 V_4 - C_4 C_4 - S_4 S_4 |^2 \, \Lambda^4 \ . 
\end{equation}
One can now add the standard Klein bottle projection
\begin{equation}
{\cal K} = \frac{1}{2} \, (V_8 - S_8 ) \, P^4  =
\frac{1}{2} \, (V_4 O_4 + O_4 V_4 - C_4 C_4 - S_4 S_4 )\, P^4 \ ,
\end{equation}
where, for later convenience, we have explicitly decomposed the SO(8) 
characters into products of SO(4) ones. 
The transverse-channel annulus amplitude
\begin{eqnarray}
\tilde{\cal A} &=& \frac{2^{-5}}{2} \Biggl\{ 
(V_4 O_4 + O_4 V_4 - C_4 C_4 - S_4 S_4  ) 
\left[
(N_{+}^{2} + N_{-}^{2} ) v W^4 + (D_{+}^{2} + D_{-}^{2} ) \frac{1}{v}
P^4 \right] 
\nonumber \\
&+& 2\, (V_4 O_4 + O_4 V_4 + C_4 C_4 + S_4 S_4 ) \left[ N_+ N_-
\, v\,  W^3 (-)^n W + D_+ D_- \, \frac{1}{v}\, P^3 (-)^m P \right] 
\nonumber \\
&+& 2\, (V_4 O_4 - O_4 V_4 - C_4 C_4 + S_4 S_4 )
 \left( 
\frac{2\eta}{\vartheta_2}\right)^2 (N_+ D_+ \, + \, N_- D_- ) 
\nonumber \\
&+& 2\, (V_4 O_4 - O_4 V_4 + C_4 C_4 - S_4 S_4 )
\left( 
\frac{2\eta}{\vartheta_2}\right)^2 (N_+ D_- \, + \, N_- D_+ ) \Biggr\}
\end{eqnarray}
is somewhat unconventional, and thus
deserves some comments. Actually, the 99, $\bar 9 \bar 9$, 
$9\bar 9$, 55, $\bar 5 \bar 5$ and $5\bar 5$ contributions have 
already been discussed previously and do not need further
explanations, but one should notice the presence of Wilson lines 
(brane displacements) in the $9\bar 9$
($5\bar 5$) sectors. These are to affect {\it different directions}, 
and are needed to lift the tachyons that would 
otherwise be present in the 
open spectrum. The transverse-channel amplitude $\tilde{\cal A}$, however,
includes additional mixed terms, absent in ordinary toroidal 
constructions, that are
to be interpreted as orbifold-like projections. In fact, the
simultaneous presence of 9 and 5-branes, a familiar feature of 
orbifold models, halves 
the number of supersymmetries and results in the corresponding 
($Z_2$) breaking of the
characters $V_8$ and $S_8$. Of course, due to the simultaneous 
presence of branes
and antibranes, the complete theory is not supersymmetric, as implied by
the different signs in the RR sectors. 
The orbifold-like structure of the annulus 
amplitude can be better appreciated after an $S$ transformation to the 
direct-channel amplitude
\begin{eqnarray}
{\cal A} &=& \frac{1}{2} (V_4 O_4 + O_4 V_4 - C_4 C_4 - S_4 S_4) 
\left[ (N_{+}^{2} + N_{-}^{2} )
\, P^4 + (D^{2}_{+} + D^{2}_{-})\, W^4 \right] 
\nonumber \\
&+& (O_4 O_4 + V_4 V_4 - S_4 C_4 - C_4 S_4 ) \left[ N_+ N_- \, P^3
P^{(1/2)} + D_+ D_- \, W^3 W^{(1/2)} \right] 
\nonumber \\
&+& (O_4 C_4 - S_4 O_4 + V_4 S_4 - C_4 V_4)
 \left( \frac{\eta}{\vartheta_4}\right)^2 \, (N_+ D_+ + N_- D_- ) 
\nonumber \\
&+& (O_4 S_4 - C_4 O_4 + V_4 C_4 - S_4 V_4) 
\left( \frac{\eta}{\vartheta_4}\right)^2 \, (N_+ D_- + N_- D_+ ) \ .
\end{eqnarray}
While the mixed $ND$ terms have the structure familiar from
$Z_2$ orbifolds, no breaking terms are actually present in this 
case both for $N$ and $D$ 
charges. This is precisely as demanded by the
consistency conditions for open-string constructions, since
only the toroidal bulk states are allowed to propagate in the 
transverse channel.

One can now derive the transverse-channel M{\"o}bius amplitude combining
the terms in $\tilde{\cal K}$ and $\tilde{\cal A}$ at the origin of the
lattices, so that
\begin{eqnarray}
\tilde{\cal M}_0 &=& \frac{2}{2} \sqrt{v} \Biggl\{
\epsilon \hat V_4  \hat O_4 \left[ (N_+
+ N_- ) \sqrt{v} + (D_+ + D_- ) \frac{1}{\sqrt{v}} \right] 
\nonumber \\
&+& \epsilon \hat O_4  \hat V_4 \left[ (N_+
+ N_- ) \sqrt{v} - (D_+ + D_- ) \frac{1}{\sqrt{v}} \right]
\nonumber \\
&+& \hat C_4  \hat C_4 \left[ (N_+
- N_- ) \sqrt{v} + (D_+ - D_- ) \frac{1}{\sqrt{v}} \right] 
\nonumber \\
&+& \hat S_4  \hat S_4 \left[ (N_+
- N_- ) \sqrt{v} - (D_+ - D_- ) \frac{1}{\sqrt{v}} \right] \Biggr\} \quad ,
\label{tmbt}
\end{eqnarray}
and extract two NS-NS and two RR tadpole conditions for this 
class of models:
\begin{eqnarray}
V_4 O_4\, &:& \quad (N_+ + N_- + 32\epsilon ) \sqrt{v} + \frac{(D_+ +
D_-)}{\sqrt{v}} \, = 0 \ ,
\nonumber \\
O_4 V_4\, &:& \quad (N_+ + N_- + 32\epsilon ) \sqrt{v} - \frac{(D_+ +
D_-)}{\sqrt{v}} \, = 0 \ ,
\nonumber \\
C_4 C_4\, &:& \quad (N_+ - N_- - 32) \sqrt{v} + \frac{(D_+ -
D_-)}{\sqrt{v}} \, = 0 \ ,
\nonumber \\
S_4 S_4\, &:& \quad (N_+ - N_- - 32 ) \sqrt{v} - \frac{(D_+ -
D_-)}{\sqrt{v}} \, = 0 \ .
\end{eqnarray}
The RR tadpole conditions demand that the net charge of the 
${\rm D}9/{\rm D}\bar 9$-branes,  $N_{+}-N_{-}$,
be equal to 32, and that the numbers of D5 and D$\bar 5$ branes
be the same. It is then impossible to satisfy the NS-NS tadpoles, and as
a result a potential is generated for the six-dimensional dilaton, 
$\phi_6$, and for 
the volume $v$ of the internal torus:
\begin{equation}
V_{{\rm eff}} 
\sim e^{-\phi_6} \left[ (N_+ + N_- + 32\epsilon ) \sqrt{v} + \frac{(D_+ +
D_-)}{\sqrt{v}} \right] \ .
\end{equation}
This potential has the peculiar property of stabilizing the internal 
volume to the value
\begin{equation}
v_{{0}} = \frac{D_+ + D_-}{N_+ + N_- + 32\epsilon} \ , 
\end{equation}
while giving a mass to the corresponding (breathing-mode) field.
This feature is common to all
models with 9, $\bar 9$, 5 and $\bar 5$ branes that we shall
encounter in the following Sections.

The direct-channel M{\"o}bius amplitude
\begin{eqnarray}
{\cal M} &=& \frac{1}{2} \Biggl\{ \epsilon \, (N_+ + N_- ) \, (\hat V
_4 \hat O_4 + \hat O_4 \hat V_4 ) \, P^4 - \epsilon \, (D_+ + D_- )\, 
(\hat V _4 \hat O_4 - \hat O_4 \hat V_4 ) \, \left( 
\frac{2\hat\eta}{\hat\vartheta_2}\right)^2 
\nonumber \\
&+& (N_+ - N_- ) \, (\hat C_4 \hat C_4 + \hat S_4 \hat S_4 )\, P^4 -
(D_+ - D_- )\, (\hat C_4 \hat C_4 - \hat S_4 \hat S_4 )\,
\left( \frac{2\hat\eta}{\hat\vartheta_2}\right)^2 \Biggr\} 
\end{eqnarray}
follows from (\ref{tmbt}) after one includes the massive contributions
and performs a $P$ transformation. The massless
open spectrum depends on the sign $\epsilon$, that is not
fixed by tadpole conditions. For $\epsilon =-1$, it comprises gauge 
bosons in the adjoint of
${\rm SO} (N_+ )_9$ $\times$ 
${\rm SO} (N_-)_{\bar 9}$ 
$\times$ ${\rm USp} (D_+)_5$ $\times$ 
${\rm USp} (D_- )_{\bar 5}$, with the matter
\begin{eqnarray}
4\ {\rm scalars} &:& (A,1;1,1) \ , \quad 
(1,A;1,1) \ , \quad (1,1;A,1) \ , \quad (1,1;1,A) \ ,
\nonumber \\
2\ {\rm scalars} &:& (f,1;f,1) \ , \quad 
(1,f;1,f) \ , \quad (1,f;f,1) \ , \quad (f,1;1,f) \ ,
\nonumber \\
\hbox{\rm L-spinor} &:& (A,1;1,1) \ , \quad 
(1,S;1,1) \ , \quad (1,1;S,1) \ , \quad (1,1;1,A) \ ,
\nonumber \\
\hbox{\rm R-spinor} &:& (A,1;1,1) \ , \quad 
(1,S;1,1) \ , \quad (1,1;A,1) \ , \quad (1,1;1,S) \ ,
\nonumber \\
\hbox{\rm half\ L-spinor} &:& (1,f;f,1) \ , \quad 
(f,1;1,f) \ ,
\nonumber \\
\hbox{\rm half\ R-spinor} &:& (f,1;f,1) \ , \quad 
(1,f;1,f)  \ ,
\nonumber
\end{eqnarray}
where $S$ ($A$) and $f$ denote the (anti)symmetric and fundamental 
representations.
On the other hand, for $\epsilon =+1$ symmetric and antisymmetric
representations are interchanged in the bosonic sector (thus 
also interchanging orthogonal and symplectic gauge groups), 
while the massless closed states, still 
supersymmetric, fill the ${\cal N}=(1,1)$ supergravity multiplet. 
It should be appreciated that, to lowest order, after a toroidal
compactification to four dimensions, for $\epsilon=+1$ $(\epsilon=-1)$
the spectrum on the D9 and D5 (D$\bar{9}$ and D$\bar{5}$) branes has
${\cal N}=2$ supersymmetry, if suitable gauge subgroups for the
D$\bar{9}$ and D$\bar{5}$ (D9 and D5) branes are regarded as global
symmetries. Thus, to lowest order the resulting massless spectrum
has ${\cal N}=4$ supersymmetry in the bulk,  ${\cal N}=2$ supersymmetry
on the branes (antibranes), while it is not supersymmetric on the
antibranes (branes).
The open sector is {\it chiral} and leads to residual mixed and gauge
anomalies, summarized by the polynomial
\begin{equation}
{\cal I}_8 = \frac{1}{8} \left( {\rm tr} R^2 - {\rm tr} F_{N_+}^{2} + {\rm
tr} F_{N_-}^{2} \right) \left( {\rm tr} F^{2}_{D_+} - 
{\rm tr} F^{2}_{D_-} \right) \ ,
\nonumber
\end{equation}
and removed by Green-Schwarz couplings \cite{gs} 
involving the single antisymmetric tensor present in the massless
closed spectrum,
while the RR tadpole conditions guarantee the cancellation of all
irreducible ${\rm tr} R^4$ and ${\rm tr} F^4$ anomalies. 

One can further deform these spectra, allowing for a
non-trivial quantized NS-NS $B_{ij}$ of rank $r$. Due to the peculiar 
structure of the models, the
closed sector behaves as in standard toroidal compactifications
\cite{toroidal}, whereas the annulus amplitude, modified as in
\cite{comm}, in the transverse channel reads
\begin{eqnarray}
\tilde{\cal A} &=& \frac{2^{-5}}{2} \Biggl\{ (V_4 O_4 + O_4 V_4 - C_4
C_4 - S_4 S_4 ) \Biggl[ 2^{r-4} ( N_+^2 + N_-^2 ) v \sum_{\{\varphi_j \}
=\pm 1} 
\sum_{\{n_i\}} W_{n_i} e^{i\pi n_i B_{ij} \varphi_j} 
\nonumber
\\
& &+ (D_+^2 + D_-^2)
\frac{1}{v} \sum_{\{m_i\}} P_{m_i} \Biggr]
\nonumber 
\\
& &+ 2 (V_4 O_4 + O_4 V_4 + C_4
C_4 + S_4 S_4 ) \Biggl[ 2^{r-4} \, N_+ N_- v \sum_{\{\varphi_j\} =\pm 1} 
\sum_{\{n_i\}} (-1)^{n_4} W_{n_i} e^{i\pi n_i B_{ij} \varphi_j} 
\nonumber 
\\
& &+ D_+ D_-
\frac{1}{v} \sum_{\{m_i\}} (-1)^{m_3} P_{m_i} \Biggr]
\nonumber
\\
& &+ 2 \times 2^{r/2}\, (V_4 O_4 - O_4 V_4 - C_4 C_4 + S_4 S_4 )
\left( \frac{2\eta}{\vartheta_2}\right)^2 ( N_+ D_+ + N_- D_- )
\nonumber
\\
& &+ 2 \times 2^{r/2}\, (V_4 O_4 - O_4 V_4 + C_4 C_4 - S_4 S_4 )
\left( \frac{2\eta}{\vartheta_2}\right)^2 ( N_+ D_- + N_- D_+ ) \Biggr\}\ .
\end{eqnarray}
The corresponding transverse-channel M{\"o}bius amplitude
\begin{eqnarray}
\tilde{\cal M} &=& \frac{2}{2} \Biggl\{ 2^{(r-4)/2} \epsilon (N_+ +
N_- ) ( V_4 O_4 + O_4 V_4) v \sum_{\{\varphi_j\} = \pm 1} \sum_{\{n_i\}}
W_{n_i}
e^{i\pi n_i B_{ij} \varphi_j}\,\gamma_\varphi
\nonumber
\\
& &+ \epsilon (D_+ + D_- )( V_4 O_4 - O_4 V_4) \left( 
\frac{2\eta}{\vartheta_2} \right)^2
\nonumber
\\
& &+ 2^{(r-4)/2} (N_+ -
N_- ) ( C_4 C_4 + S_4 S_4) v \sum_{\{\varphi_j\} = \pm 1} \sum_{\{n_i\}}
W_{n_i} e^{i\pi n_i B_{ij} \varphi_j}\,\gamma_\varphi
\nonumber
\\
& &+ \epsilon (D_+ - D_- )( C_4 C_4 - S_4 S_4) \left( 
\frac{2\eta}{\vartheta_2} \right)^2 \Biggr\}
\end{eqnarray}
is thus fixed by factorization and, as usual, involves the signs
$\gamma_\varphi$ that enforce a proper normalization. As a result, the
net RR charge of the D9/D$\bar 9$-branes is reduced by a factor
$2^{r/2}$ and, after a modular transformation to the direct channel,
$2^{r/2}$ copies of the $ND$ sectors are present, while the 
$\gamma_\varphi$ allow for continuous interpolations between
orthogonal and symplectic gauge groups only on the D9 and D$\bar 9$ branes.


\subsection{The $T^4 /Z_2$ orbifold revisited}

Let us now reconsider the open descendants of the $T^4/Z_2$ orbifold,
allowing for the simultaneous presence of branes and antibranes of the
same type. For
the sake of clarity, let us begin from the torus amplitude
\begin{equation}
{\cal T} = \frac{1}{4} \left\{ | Q_o + Q_v |^2 \Lambda^4+
| Q_o - Q_v |^2 \left| \frac{2 \eta}{\vartheta_2} \right|^4 
+ 16 | Q_s + Q_c |^2 \left| \frac{\eta}{\vartheta_4} \right|^4 
+ 16 | Q_s - Q_c |^2 \left| \frac{\eta}{\vartheta_3} \right|^4 \right\} \ ,
\label{d1}
\end{equation}
and its Klein bottle projection
\begin{equation}
{\cal K} = \frac{1}{4} \left\{ (Q_o + Q_v ) ( P^4 + W^4 ) + 
2 \times 16  \epsilon \, (Q_s + Q_c)
\left( \frac{\eta}{\vartheta_4} \right)^2 \right\} \ ,
\end{equation}
where we have used the standard $Z_2$ decompositions 
\begin{eqnarray}
Q_o &= \ V_4 O_4 - C_4 C_4 \ , \qquad Q_v &= \ O_4 V_4 - S_4 S_4 \ , 
\nonumber \\
Q_s &= \ O_4 C_4 - S_4 O_4 \ , \qquad Q_c &= \ V_4 S_4 - C_4 V_4 \ . 
\end{eqnarray}

We have written ${\cal K}$ in this form, since one can actually 
modify the Klein bottle projection for twisted closed states, as 
in \cite{ads2}. Multiple Klein-bottle projections, first
discussed in \cite{pss}, have also interesting physical applications to 
tachyon-free non-supersymmetric open-string models \cite{notach}.
The choice $\epsilon =+1$ corresponds to the usual $T^4/Z_2$ orientifold, 
and gives a closed unoriented spectrum containing ${\cal N}=(1,0)$
supergravity coupled to 20 hypermultiplets and 1 tensor multiplet, 
while the choice $\epsilon =-1$ gives ${\cal N}=(1,0)$ supergravity 
coupled to 4 hypermultiplets and 17 tensor multiplets. 

From the Klein-bottle contribution to the massless tadpoles
\be
\tilde{\cal K}_0 = \frac{2^5}{4} \left\{
\left ( \sqrt{v}+\frac{\epsilon}{\sqrt{v}} \right )^2 Q_o
+\left ( \sqrt{v}-\frac{\epsilon}{\sqrt{v}} \right )^2 Q_v \right\} \ , 
\label{d4}
\ee
one can anticipate the need for a net number of 32 D9 and 32 (anti)D5-branes
(for $\epsilon =\pm1$, respectively) in order to cancel the RR tadpoles.

Proceeding as in the previous Sections, the annulus amplitude in the 
presence of branes and antibranes is
\begin{eqnarray}
{\cal A} &=& \frac{1}{4} \Biggl\{ 
(V_4 O_4 + O_4 V_4 - C_4 C_4 - S_4 S_4 ) 
\left[ (N_+^2 + N_-^2) \, P^4 + (D_+^2 + D_-^2) \, W^4 \right] 
\nonumber \\ 
&+& 2 ( O_4 O_4 + V_4 V_4 - S_4 C_4 - C_4 S_4) \left[ N_+ N_- \, P^3 P_{1/2} 
+ D_+ D_-  \, W^3 W_{1/2} \right] 
\nonumber \\
&+& (V_4 O_4 - O_4 V_4 - C_4 C_4 + S_4 S_4 ) 
\left( \frac{2 \eta}{\vartheta_2}\right)^2 
(R_{N_+}^2 + R_{N_-}^2 + R_{D_+}^2 + R_{D_-}^2) 
\nonumber \\
&+& 2 (O_4 C_4 + V_4 S_4 - S_4 O_4 - C_4 V_4 ) 
\left( \frac{\eta}{\vartheta_4} \right)^2
(N_+ D_+ + N_- D_-) 
\nonumber \\ 
&+& 2  (O_4 S_4 + V_4 C_4 - C_4 O_4 - S_4 V_4) \left(
\frac{\eta}{\vartheta_4} \right)^2 (N_+ D_- + N_- D_+) 
\nonumber \\
&+& 2  (O_4 C_4 - V_4 S_4 - S_4 O_4 + C_4 V_4 ) 
\left( \frac{\eta}{\vartheta_3} \right)^2 
(R_{N_+} R_{D_+} + R_{N_-} R_{D_-}) 
\nonumber \\
&+& 2  (- O_4 S_4 + V_4 C_4 - C_4 O_4 + S_4 V_4) 
\left( \frac{\eta}{\vartheta_3} \right)^2 (R_{N_+} R_{D_-} + R_{N_-} R_{D_+})
\biggr\} \ ,
\end{eqnarray}
where $N_\pm$ ($D_\pm$) refer to the D9 and 
D$\bar 9$ (D5 and D$\bar 5$) branes, respectively, and the $R$'s are
orbifold-induced breaking terms. As in the previous cases, 
in order to lift the tachyons, following
\cite{au} we have inserted  the D5 and D$\bar 5$-branes at different fixed 
points of the orbifolds, and we have added Wilson lines for the D9-branes
along a different coordinate.

Finally, the M{\"o}bius amplitude
\ba
{\cal M} &=& - \frac{1}{4} \biggl\{ (\hat V _4 \hat O_4 + \hat O_4
\hat V_4 - \hat C_4 \hat C_4 -\hat S_4 \hat S_4 ) 
\left[ N_+ \, P^4 + \epsilon D_+ \, W^4  \right] 
\nonumber \\
&+& (\hat V _4 \hat O_4 + \hat O_4
\hat V_4 + \hat C_4 \hat C_4 + \hat S_4 \hat S_4 ) 
\left[ N_- \, P^4 + \epsilon D_- \, W^4 \right] 
\nonumber \\
&-& (\hat V _4 \hat O_4 - \hat O_4
\hat V_4 - \hat C_4 \hat C_4 + \hat S_4 \hat S_4 ) 
\left( \frac{2\hat\eta}{\hat \vartheta_2 } \right)^2
( \epsilon N_+   + D_+ ) 
\nonumber \\
&-& (\hat V _4 \hat O_4 - \hat O_4
\hat V_4 + \hat C_4 \hat C_4 - \hat S_4 \hat S_4 ) 
\left( \frac{2\hat\eta}{\hat\vartheta_2} \right)^2 (\epsilon N_- +
D_- ) 
\biggr\} 
\end{eqnarray}
gives a proper symmetrization of the annulus amplitude, while
the tadpole conditions
\begin{eqnarray}
& & N_+ - N_-  = 32 \ , \qquad  D_+ - D_-  = 32 \, \epsilon \ ,
\nonumber \\ 
& & R_{N_+} \, = \, R_{N_-} \, = \, R_{D_+} \, = \, R_{D_-} = 0 
\end{eqnarray}
require net numbers of 32 
D9-branes and 32 D5-branes (antibranes) for $\epsilon =\pm 1$.

The choice $\epsilon =+1$ corresponds to a deformation of the 
supersymmetric ${\rm U} (16) \times {\rm U} (16)$ model
\cite{bs,gp}, and requires the introduction of complex charges,
so that
\begin{eqnarray}
& N_\pm = n_\pm + \bar n_\pm \ , \qquad & R_{N_\pm} = i \, ( n_\pm - \bar
n_\pm ) \ , 
\nonumber \\
& D_\pm = d_\pm + \bar d_\pm \ , \qquad & R_{D_\pm} = i \, ( d_\pm - \bar
d_\pm ) \ .
\end{eqnarray}
The resulting massless open spectrum can be extracted
from 
\begin{eqnarray}
\!\!\!\!\!\!\!\!{\cal A}_0 \!\!\!&+&\!\!\! {\cal M}_0 = (n_+ \bar n_+ 
+ n_- \bar n_- + 
d_+ \bar d_+ + d_- \bar d_- ) \, (V_4 O_4 - C_4 C_4) 
\nonumber \\
&+& \frac{1}{2} (n_+^2+\bar n _+^2 + n_-^2 + \bar n_-^2 + d_+^2 + \bar
d _+^2 + d_-^2 + \bar d_-^2) \, (O_4 V_4 - S_4 S_4 ) 
\nonumber \\
&+& (n_+ \bar d_+ + \bar n_+ d_+ + n_- \bar d_- + \bar n_- d_-) \,
(O_4 C_4 - S_4 O_4) 
\\
&-& ( n_+ \bar d_- + n_- \bar d_+ + \bar n_+ d_- + \bar n_- d_+ ) \,
 C_4 O_4 +
+ (n_+ d_- + \bar n _+ \bar d _- + n_- d_+ + \bar n _- \bar d_+ ) \,
O_4 S_4  
\nonumber \\
&-& \frac{1}{2} \Bigl\{
(n_+ + \bar n _+ + d_+ + \bar d _+) \, (\hat O_4 \hat V_4 - \hat S_4 \hat
S_4 ) 
+ (n_- + \bar n_- + d_- + \bar d_- ) \, (\hat O_4 \hat V_4 + \hat S_4 \hat
S_4 ) \Bigr\} \ .
\nonumber
\end{eqnarray}
The 99 and 55 sectors are supersymmetric, and comprise vector multiplets 
in the adjoint of 
${\rm U} (n_+)_9$ $\times$ ${\rm U} (n_-)_{\bar 9}$ 
$\times$ ${\rm U} (d_+)_5$ $\times$ 
${\rm U} (d_-)_{\bar 5}$, 
together with hypermultiplets in the antisymmetric and conjugate 
antisymmetric representations. On the other hand,  
the $\bar 9\bar 9$
and $\bar 5\bar 5$ are not supersymmetric and contain 
left-handed Weyl fermions in the antisymmetric and conjugate 
antisymmetric representations, and a quartet of real scalars 
in the symmetric and conjugate symmetric representations. 
Finally, the 95 and $\bar 9\bar 5$ sectors give hypermultiplets
in the $(n_+\,,\,\bar d _+)$ and $(n_-\,,\,\bar d _-)$, respectively,
while the $9 \bar 5$ and $\bar 9 5$ sectors, evidently
not supersymmetric, give a quartet of 
scalars in the $(n_+\,,\, d_-)$, $(n_-\,,\,d_+)$ and 
right-handed Weyl fermions in the $(n_+\,,\,\bar d _-)$,  
$(n_-\,,\,\bar d_+ )$.
The irreducible gravitational and gauge anomalies vanish as a result
of tadpole conditions, whereas the residual anomaly polynomial
\begin{equation}
{\cal I}_8 = -\frac{1}{16} \left[ {\rm tr} R^2 - 
2 ( {\rm tr} F_{N_+}^2 - {\rm tr} F_{N_-}^2) \right]
\left[ {\rm tr} R^2 - 2 ( {\rm tr} F_{D_+}^2 - {\rm
tr} F_{D_-}^2 ) \right]
\end{equation}
requires a conventional Green-Schwarz mechanism \cite{gs} involving the
single antisymmetric tensor present in the massless closed spectrum. 

The choice $\epsilon=-1$ corresponds to a deformation of the
original model in \cite{ads2}, and calls for the parametrization
\begin{eqnarray}
& N_\pm = n_\pm + m_\pm \ , \qquad & R_{N_\pm} = n_\pm - m_\pm \ , 
\nonumber \\
& D_\pm = p_\pm + q_\pm \ , \qquad & R_{D_\pm} = p_\pm - q_\pm \ ,
\end{eqnarray}
in terms of real charges. The massless open spectrum can now be read
from
\begin{eqnarray}
{\cal A}_0 \!\!\!&+&\!\!\! {\cal M}_0 = \frac{1}{2}(n_+^2 + n_-^2 
+ m_+^2 + m_-^2 + p_+^2 +
p_-^2 + q_+^2 + q_-^2)\, (V_4 O_4 - C_4 C_4 ) \nonumber
 \\
\!\!\!&+&\!\!\! (m_+ n_+ + m_- n_- + p_+ q_+ + p_- q_- )\, 
(O_4 V_4 - S_4 S_4)
\nonumber \\
\!\!\!&+&\!\!\! (m_+ q_+ + n_+ p_+ + m_- q_- + n_- p_- )\, (O_4 C_4 - S_4 O_4)
 \\
\!\!\!&+&\!\!\! (m_+ p_- + n_+ q_- + m_- p_+ + n_- q_+ )\, O_4 S_4 
- (m_+ q_- + n_+ p_- + m_- q_+ + n_- p_+ )\, C_4 O_4 \nonumber \\ 
\!\!\!&-&\!\!\! \frac{1}{2} \Bigl\{ 
(n_+ + m_+ - p_+ - q_+ ) (\hat V_4 \hat O_4 \!-\! \hat C_4 \hat C_4 ) \!+\!
(n_- + m_- - p_- - q_- ) (\hat V_4 \hat O_4 \!+\! \hat C_4 \hat C_4 )
\Bigr\} \ , \nonumber
\end{eqnarray}
and the massless excitations thus comprise gauge bosons in the adjoint of
${\rm SO} (m_+)_9$ $\times$ ${\rm SO} (n_+)_9$
$\times$
${\rm SO} (m_-)_ {\bar 9}$ $\times$ ${\rm SO} (n_-)_{\bar 9}$ 
$\times$ ${\rm USp} (p_+)_5$ $\times$ ${\rm USp} (q_+)_5$ $\times$ 
${\rm USp} (p_-)_{\bar 5}$ $\times$ ${\rm USp} (q_-)_{\bar 5}$,
left-handed Weyl fermions in the adjoint of ${\rm SO}
(m_+),$ ${\rm SO} (n_+),$ ${\rm USp} (p_+)$ and ${\rm USp} (q_+),$ in
the symmetric representations of ${\rm SO} (m_-)$ and ${\rm SO}
(n_-)$ and in the antisymmetric representations of 
${\rm USp} (p_-)$ and ${\rm USp} (q_-).$ They also comprise a full
hypermultiplet in the $(m_+\,,\, n_+),$ $(m_-\,,\, n_-),$ 
$(p_+ \,,\, q_+),$ $( p_-\,,\, q_- ),$ as well as half
hypermultiplets in the $(m_+\,,\, q_+),$ $(n_+\,,\, p_+),$
$(m_- \,,\, q_-)$ and $(n_- \,,\,p_- ).$ Finally, the $9\bar 5$ and
$5\bar 9$ sectors comprise complex scalars in the $(m_+\,,\, p_-),$
$(n_+\,,\, q_-),$ $(m_-\,,\, p_+)$ and $(n_- \,,\,q_+ ),$ and 
left-handed symplectic Majorana-Weyl spinors in
the $(m_+\,,\, q_-),$ $(n_+ \,,\,p_-),$ $(m_-\,,\, q_+)$
and $(n_- \,,\,p_+).$ 

The tadpole conditions eliminate all 
irreducible gauge and gravitational anomalies, whereas the 
residual reducible anomaly
\begin{eqnarray}
{\cal I}_8 &=& \frac{1}{64} 
\left[ 2 {\rm tr} R^2 - \left( {\rm tr} F_{m_+}^2+
{\rm tr} F_{n_+}^2 - {\rm tr} F_{m_-}^2 - {\rm tr} F_{n_-}^2 - 
{\rm tr} F_{p_+}^2 - {\rm tr} F_{q_+}^2 + {\rm tr} F_{p_-}^2
+ {\rm tr} F_{q_-}^2 \right) \right]^2 
\nonumber 
\\
&-& \frac{1}{64} \left( 
{\rm tr} F_{m_+}^2 + {\rm tr} F_{n_+}^2
- {\rm tr} F_{m_-}^2 - {\rm tr} F_{n_-}^2 + {\rm tr} F_{p_+}^2 + 
{\rm tr} F_{q_+}^2 - {\rm tr} F_{p_-}^2 - {\rm tr} F_{q_-}^2 \right)^2
\nonumber 
\\ 
&+& \frac{1}{64} \left( {\rm tr} F_{m_+}^2 - {\rm tr} F_{n_+}^2
- {\rm tr} F_{m_-}^2 + {\rm tr} F_{n_-}^2 - {\rm tr} F_{p_+}^2 + 
{\rm tr} F_{q_+}^2 + {\rm tr} F_{p_-}^2 - {\rm tr} F_{q_-}^2
\right)^2 
\nonumber 
\\
&+& \frac{4}{64} \left( {\rm tr} F_{m_+}^2 - {\rm tr} F_{n_+}^2
+ {\rm tr} F_{m_-}^2 - {\rm tr} F_{n_-}^2 \right)^2 
\nonumber 
\\
&+& \frac{4}{64} \left( {\rm tr} F_{p_+}^2 - {\rm tr} F_{q_+}^2 
+ {\rm tr} F_{p_-}^2 - {\rm tr} F_{q_-}^2 \right)^2 
\nonumber 
\\
&+& \frac{3}{64} \left( {\rm tr} F_{m_+}^2 - {\rm tr} F_{n_+}^2
- {\rm tr} F_{m_-}^2 + {\rm tr} F_{n_-}^2 + {\rm tr} F_{p_+}^2
- {\rm tr} F_{q_+}^2 - {\rm tr} F_{p_-}^2 + {\rm tr} F_{q_-}^2
\right)^2 
\end{eqnarray}
can be removed 
by a generalized Green-Schwarz mechanism \cite{ggs}.

In both models a potential is generated for the six-dimensional dilaton and
the volume of the internal manifold
\begin{equation}
V_{{\rm eff}} 
\sim e^{-\phi_6} \left[ (N_+ + N_- - 32) \sqrt{v} + \frac{(D_+ +
D_- - 32 \e )}{\sqrt{v}} \right] \ , 
\end{equation}
as a result of uncancelled NS-NS tadpoles.  As in the
previous model, this stabilizes the internal volume at a
local minimum, that in this case is
\begin{equation}
v_{0} = \frac{D_+ + D_- -32 \e }{N_+ + N_- - 32} \ , 
\end{equation}
and gives mass to the corresponding (breathing-mode) field.

In the present models discrete deformations for
the NS-NS $B$-field may be introduced as in
\cite{toroidal,comm}, and result in a reduced total RR charge for nine and
five (anti)branes and in multiplicities for the $ND$ sectors. Moreover,
the corresponding signs in the M{\"o}bius amplitude allow continuous 
interpolations between orthogonal (symplectic) and unitary gauge groups.

\subsection{The non-supersymmetric $T^6/Z_4$ orbifold revisited}

As a last example, we would like to generalize the non-supersymmetric 
$T^6/Z_4$ orbifold of Section 3, allowing for the simultaneous 
presence of branes and
antibranes. As in the previous examples, the torus and Klein 
bottle amplitudes are not affected, whereas the annulus and M{\"o}bius 
amplitudes are now given by
\begin{eqnarray}
{\cal A} \!\!&=& \!\! \frac{1}{8} \Biggl\{ 
\left[ ( N_+^2 + N_-^2 ) P_1 P_2 P_3 + ( D_+^2 + D_-^2 ) W_1 W_2 P_3
\right] \psi_{00}
\nonumber 
\\
\!\!&+&\!\! ( R_{N_+}^2 + R^2_{N_-} + R_{D_+}^2 + R_{D_-}^2 ) \psi_{01}
\phi_{01}^2 \left( \frac{2\eta}{\vartheta_2} \right) 
+ ( S_{N_+}^2 + S^2_{N_-} + S_{D_+}^2 + S_{D_-}^2 )
\psi_{02} \left( \frac{2\eta}{\vartheta_2} \right)^2 P_3 
\nonumber 
\\
\!\!&+&\!\! ( T_{N_+}^2 + T^2_{N_-} + T_{D_+}^2 + T_{D_-}^2 ) \psi_{03}
\phi_{03}^2 \left( \frac{2\eta}{\vartheta_2} \right) 
+ 2 ( N_+ N_- P_1 P^{1/2}_2 P_3 + D_+D_- W_1 
W_2^{1/2} P_3 ) \psi_{00}^{(-)} 
\nonumber 
\\
\!\!&+&\!\! 2 ( N_+ D_+ + N_- D_- ) \psi_{20}
\left(\frac{\eta}{\vartheta_4} \right)^2 P_3 +
2 ( R_{N_+} R_{D_+} + R_{N_-} R_{D_-} ) \psi_{21}
\phi_{21}^2 \left( \frac{2\eta}{\vartheta_2} \right) 
\\ 
\!\!&+&\!\! 2 ( S_{N_+} S_{D_+} + R_{N_-} R_{D_-} ) \psi_{22}
\left( \frac{\eta}{\vartheta_3} \right)^2 P_3 +
2 ( T_{N_+} T_{D_+} + R_{N_-} R_{D_-} ) \psi_{23}
\phi_{23}^2 \left( \frac{2\eta}{\vartheta_2} \right) 
\nonumber 
\\
\!\!&+&\!\! 2 ( N_+ D_- + N_- D_+ ) \psi^{(-)}_{20}
\left(\frac{\eta}{\vartheta_4} \right)^2 P_3 +
2 ( R_{N_+} R_{D_-} + R_{N_-} R_{D_+} ) \psi^{(-)}_{21}
\phi_{21}^2 \left( \frac{2\eta}{\vartheta_2} \right) 
\nonumber 
\\ 
\!\!&+&\!\! 2 ( S_{N_+} S_{D_-} + R_{N_-} R_{D_+} ) \psi^{(-)}_{22}
\left( \frac{\eta}{\vartheta_3} \right)^2 P_3 +
2 ( T_{N_+} T_{D_-} + R_{N_-} R_{D_+} ) \psi^{(-)}_{23}
\phi_{23}^2 \left( \frac{2\eta}{\vartheta_2} \right)
\Biggr\} 
\nonumber
\end{eqnarray}
and
\begin{eqnarray}
{\cal M} &=& - \frac{1}{8} \Biggl\{ 
( N_+ P_1 P_2 P_3 - D_+ W_1 W_2 P_3 )
\hat\psi_{00} + ( N_+ - D_+) P_3 \hat\psi_{02}
\left( \frac{2\hat\eta}{\hat\vartheta_2}\right )^2 
\nonumber 
\\
&+& ( N_- P_1 P_2 P_3 - D_- W_1 W_2 P_3) \hat{\tilde\psi}{}_{00}^{(-)} 
+ ( N_- - D_- ) P_3 \hat{\tilde\psi}{}_{02}^{(-)} 
\left( \frac{2\hat\eta}{\hat\vartheta_2} \right )^2 
\nonumber 
\\
&+& 2 ( S_{N_+} - S_{D_+} ) \hat\psi_{01} \hat\phi_{01}^2
\left( \frac{2\hat\eta}{\hat\vartheta_2} \right) 
+ 2 ( S_{N_+} - S_{D_+} ) \hat\psi_{03} \hat\phi_{03}^2
\left( \frac{2\hat\eta}{\hat\vartheta_2} \right)
\nonumber \\
&+& 2 ( S_{N_-} - S_{D_-} ) \hat{\tilde\psi}{}_{01}^{(-)} 
\hat\phi_{01}^2
\left( \frac{2\hat\eta}{\hat\vartheta_2} \right) 
+ 2 ( S_{N_-} - S_{D_-} ) \hat{\tilde\psi}{}_{03}^{(-)} 
\hat\phi_{03}^2
\left( \frac{2\hat\eta}{\hat\vartheta_2} \right) 
\Biggr\} \ , 
\end{eqnarray}
where the $\psi$'s and $\phi$'s have been introduced in Section 3,
$P^{1/2}$ and $W^{1/2}$ denote shifted momentum and winding sums, and
we are parametrizing again the 9, $\bar 9$, 5 and $\bar
5$-brane charges by $N_\pm$, $D_\pm$ and by their orbifold-induced
breakings $R$, $S$ and $T$. 
The $Z_4$ action suggests the parametrization
\begin{eqnarray}
& N_\pm = n_\pm + m_\pm + p_\pm + \bar m _\pm \ , 
& D_\pm = q_\pm + r_\pm + s_\pm + \bar r _\pm \ , 
\nonumber \\
& R_{N_\pm} = n_\pm + i m_\pm - p_\pm - i \bar m_\pm \ ,
& R_{D_\pm} = q_\pm + i r_\pm - s_\pm - i \bar r_\pm \ , 
\nonumber \\ 
& S_{N_\pm} = n_\pm - m_\pm + p_\pm - \bar m_\pm \ , 
& S_{D_\pm} = q_\pm - r_\pm + s_\pm - \bar r_\pm \ ,
\nonumber \\ 
& T_{N_\pm} = n_\pm - i m_\pm - p_\pm + i \bar m_\pm \ , 
& T_{D_\pm} = q_\pm - i r_\pm - s_\pm + i \bar r_\pm 
\end{eqnarray}
and, as expected from the simpler case discussed in Section 3, the RR tadpole
conditions fix to 32 the net numbers of D9-branes and D$\bar
5$-branes, and require the vanishing of all the $R$, $S$ and $T$ breaking
coefficients.
Even in this case the NS-NS tadpole conditions can not be satisfied,
and as a result the potential 
\begin{equation}
V_{{\rm eff}} 
\sim e^{-\phi_4}\sqrt{v_3} \left[ (N_+ + N_- - 32) \sqrt{v_1v_2} + \frac{(D_+ +
D_- + 32)}{\sqrt{v_1v_2}} \right] 
\end{equation}
is generated
for the four-dimensional dilaton and the volume of the
internal manifold. As in the previous cases, it constrains
the internal volumes, since it has a minimum for
\begin{equation}
(v_1v_2)_{{0}} = \frac{D_+ + D_- +32}{N_+ + N_- - 32} \ .
\end{equation}

From the tadpole conditions and the massless contributions to 
${\cal A}$ and ${\cal M}$, one finds the Chan-Paton gauge group
${\rm SO} (n_+)_9$ $\times$ ${\rm SO} (p_+)_9$ $\times$ ${\rm U} 
(m_+)_9$ $\times$ ${\rm SO} (n_-) _{\bar 9}$ $\times$ 
${\rm SO} (p_-)_{\bar 9}$ $\times$ 
${\rm U} (m_-)_{\bar 9}$ $\times$
${\rm USp} (q_+)_5$ $\times$ ${\rm USp} (s_+)_5$ $\times$ ${\rm U}
(r_+)_5$ 
$\times$ ${\rm USp} (q_-)_{\bar 5}$ $\times$ ${\rm USp} (s_-)_{\bar 5}$
$\times$ ${\rm U} (r_-)_{\bar 5}$.
The 99 and 55 sectors are supersymmetric and, aside from ${\cal N}=1$ vector
multiplets in the adjoint of the gauge group, comprise
chiral multiplets in the representations\footnote{For the sake of
brevity, we denote by $S_\alpha$ ($A_\alpha$)
the (anti)symmetric
representations for the $\alpha$-th factor in the gauge group, by
$\bar{S}_\alpha$ ($\bar{A}_\alpha$) their conjugates,
and by Adj the adjoint of unitary factors.}
\begin{eqnarray}
99: & & 2 (n_+,\bar m _+)\ , \quad 2 (p_+,m_+) \ , \quad  (n_+ , p_+) \ , \quad
A_{m_+}\ , \quad\bar A_{m_+}  \ , 
\nonumber \\
55: & & 2 (q_+,\bar r_+)\ , \quad 2 (s_+,r_+) \ , \quad  (q_+ , s_+) 
\ , \quad S_{r_+} \ , \quad \bar S _{r_+} \ .
\nonumber
\end{eqnarray}
The $\bar 9 \bar 9$ and $\bar 5 \bar 5$ sectors are non-supersymmetric
and, aside from the gauge bosons in the adjoint of the
corresponding gauge groups, comprise Dirac spinors 
in the representations
\begin{eqnarray}
\bar 9 \bar 9: & & S_{n_-} \ , \quad S_{p_-} \ , \quad {\rm Adj}_{m_-}
\ , \quad S_{m_-} \ , \quad\bar S _{m_-} \ , 
\nonumber \\
\bar 5 \bar 5: & & A_{q_-} \ , \quad A_{s_-} \ , \quad  
{\rm Adj}_{r_-} \ , \quad A_{r_-} \ , \quad \bar A_{r_-} 
\ ,
\nonumber
\end{eqnarray}
scalars in the representations $S_{m_-}$ and $\bar{S}_{m_-}$ 
($A_{r_-}$ and $\bar{A}_{r_-}$),
pairs of chiral multiplets in the representations
$(n_-,\bar{m}_-)$ and $(p_-,m_-)$ ($(q_-,\bar{r}_-)$ and $(s_-,r_-)$),
and a single chiral multiplet in the representation $(n_-,p_-)$ ($(q_-,s_-)$)
for the ${\bar 9}{\bar 9}$ (${\bar 5}{\bar 5}$) sectors.
The $9\bar{5}$ spectrum contains scalars in the representations
$(n_+,q_-)$, $(p_+,s_-)$, $(m_+,\bar{r}_-)$ and $(\bar{m}_+,r_-)$, 
and spinors in the representations 
$(n_+,\bar{r}_-)$, $(p_+,r_-)$, $(m_+, s_-)$ and $(\bar{m}_+,q_-)$,
while the 59 sector contains chiral multiplets
in the representations  $(n_+,\bar{r}_+)$, $(p_+,r_+)$, 
$(m_+,s_+)$ and $(\bar{m}_+,q_+)$. Finally, the 5$\bar{9}$ and 
$\bar{5}\bar{9}$ sectors give similar
contributions, with suitable relabelings of the Chan-Paton charges.
This chiral spectrum is free of non-abelian gauge anomalies.


\section{Conclusions}

In this paper we have discussed type I models 
where the basic two-dimensional consistency conditions (RR tadpoles),
although apparently unsolvable, can actually be solved provided
supersymmetry is  broken {\it at the string scale} 
on some (anti)branes, while some amount of supersymmetry is left,
to lowest order, on other branes, and most
notably in the bulk (gravitational) sector.

This new feature of our mechanism (called  in \cite{ads2} brane 
supersymmetry breaking), where the breaking is tied to the 
{\it string scale}, is the main difference compared to 
other previously known mechanisms in heterotic strings
\cite{closed}, in type I strings 
\cite{ads,adds2,bachas,bd,aaf,bkl} or in M-theory \cite{aq}. 
Brane supersymmetry breaking
is typically accompanied by uncanceled NS-NS tadpoles for the dilaton 
at some moduli fields of the compact manifold.

Since our models contain non-BPS brane-antibrane systems,
one might wonder whether they correspond to stable configurations.
This is indeed the case, and actually the ``minimal'' 
D9-D${\bar 5}$ pairs of \cite{ads2}, aside from having no 
open-string tachyons, experience no net mutual forces, as can be
seen from the vanishing of the corresponding annulus amplitude. 
Following \cite{su,au}, these models can be further deformed by the 
inclusion of additional brane-antibrane pairs of the same type
(D9-D${\bar 9}$, D$5_i$-D${\bar 5}_i$). We would like to stress that this 
additional deformation is an interesting option, not required 
by any RR tadpole conditions. Rather, it destroys local tadpole cancellations,
while the new brane-antibrane pairs 
do experience mutual forces, although the tachyons resulting from strings
stretched between them
may be lifted if the endpoints are suitably separated in the internal space. 
Interestingly, the NS-NS tadpoles resulting from the various mutual
interactions between (anti)branes and orientifold planes 
determine a scalar potential that can actually
stabilize some of the radii of the compact space, giving masses to
the corresponding fields. Large (small) volumes, however, require 
unnaturally large 
numbers of D5-D$\bar{5}$ (D9-D$\bar{9}$) pairs, whose presence, therefore,
asks for a dynamical reason. Since
the whole approach is perturbative, the dilaton potential has, not
surprisingly, a runaway behavior towards vanishing string
coupling. We would like to stress that all these constructions, despite their
attractive features, do not solve the problem of the (four-dimensional)
cosmological constant.

Breaking supersymmetry at the string scale is a viable phenomenological
alternative if our world is a non-supersymmetric (anti)brane
and the string scale $M_I$ is in the TeV range, or if our world is on
a brane that, to lowest order, is supersymmetric. In 
the latter case, supersymmetry breaking could
be mediated by gauge interactions, and a realistic spectrum would ask for
a string scale of the order of 100 TeV if the non-supersymmetric
(anti)brane gauge coupling were of order one, or at an intermediate value
if the (anti)brane gauge coupling were suppressed
by the volume of the internal space. Alternatively, if this coupling were
very tiny, gravitation would mediate supersymmetry breaking with 
an intermediate string scale of the order of $10^{11}$  GeV \cite{int}.  
The (tree-level) supersymmetry present in the bulk sector can have
far-reaching physical consequences. Indeed, quantum corrections to
brane/antibrane couplings can also be interpreted in terms of
effective brane
couplings to bulk fields. If these respect the bulk symmetry,
one could contemplate the fascinating possibility of living in a
non-supersymmetric world where quantum corrections are governed by a
supersymmetric bulk sector. In this case, the 
gauge hierarchy would be protected and the quantum corrections would be 
very similar to those of a supersymmetric brane theory. 
We will return to these interesting issues in the near future.


\vfill\eject
\begin{flushleft}
{\large \bf Acknowledgments}
\end{flushleft}
C.A. and A.S. would like to thank the Physics Department of the 
Humboldt University, 
C.A. and E.D. would like thank the Physics Department of
the University of Rome ``Tor Vergata'', and A.S. would like to thank the
Centre de Physique Th{\'e}orique of the {\'E}cole Polytechnique 
for the warm hospitality extended
to them while this work was in progress. 
This research was supported in part by the EEC under TMR contract 
ERBFMRX-CT96-0090, and in part by the National Science Foundation, under
Grant No. PHY94-07194.

\vfill\eject

\appendix
\section{Characters for the $T^6/Z_2 \times Z_2$ orbifolds}

In this Appendix we list the $Z_2 \times Z_2$ 
characters needed for the models in Section 2. They may be
expressed as ordered products of the four SO(2) level-one 
characters, $O_2$, $V_2$, $S_2$ and $C_2$:

\ba
\tau_{oo}&=&V_2O_2O_2O_2+O_2V_2V_2V_2-S_2S_2S_2S_2-C_2C_2C_2C_2 \ , 
\nonumber \\
\tau_{og}&=&O_2V_2O_2O_2+V_2O_2V_2V_2-C_2C_2S_2S_2-S_2S_2C_2C_2 \ , 
\nonumber \\
\tau_{oh}&=&O_2O_2O_2V_2+V_2V_2V_2O_2-C_2S_2S_2C_2-S_2C_2C_2S_2 \ , 
\nonumber \\
\tau_{of}&=&O_2O_2V_2O_2+V_2V_2O_2V_2-C_2S_2C_2S_2-S_2C_2S_2C_2 \ , 
\nonumber \\
\tau_{go}&=&V_2O_2S_2C_2+O_2V_2C_2S_2-S_2S_2V_2O_2-C_2C_2O_2V_2 \ , 
\nonumber \\
\tau_{gg}&=&O_2V_2S_2C_2+V_2O_2C_2S_2-S_2S_2O_2V_2-C_2C_2V_2O_2 \ , 
\nonumber \\
\tau_{gh}&=&O_2O_2S_2S_2+V_2V_2C_2C_2-C_2S_2V_2V_2-S_2C_2O_2O_2 \ , 
\nonumber \\
\tau_{gf}&=&O_2O_2C_2C_2+V_2V_2S_2S_2-S_2C_2V_2V_2-C_2S_2O_2O_2 \ , 
\nonumber \\
\tau_{ho}&=&V_2S_2C_2O_2+O_2C_2S_2V_2-C_2O_2V_2C_2-S_2V_2O_2S_2 \ , 
\nonumber \\
\tau_{hg}&=&O_2C_2C_2O_2+V_2S_2S_2V_2-C_2O_2O_2S_2-S_2V_2V_2C_2 \ , 
\nonumber \\
\tau_{hh}&=&O_2S_2C_2V_2+V_2C_2S_2O_2-S_2O_2V_2S_2-C_2V_2O_2C_2 \ , 
\nonumber \\
\tau_{hf}&=&O_2S_2S_2O_2+V_2C_2C_2V_2-C_2V_2V_2S_2-S_2O_2O_2C_2 \ , 
\nonumber \\
\tau_{fo}&=&V_2S_2O_2C_2+O_2C_2V_2S_2-S_2V_2S_2O_2-C_2O_2C_2V_2 \ , 
\nonumber \\
\tau_{fg}&=&O_2C_2O_2C_2+V_2S_2V_2S_2-C_2O_2S_2O_2-S_2V_2C_2V_2 \ , 
\nonumber \\
\tau_{fh}&=&O_2S_2O_2S_2+V_2C_2V_2C_2-C_2V_2S_2V_2-S_2O_2C_2O_2 \ , 
\nonumber \\
\tau_{ff}&=&O_2S_2V_2C_2+V_2C_2O_2S_2-C_2V_2C_2O_2-S_2O_2S_2V_2 \ . 
\label{a3}
\ea
While these are sufficient to describe all supersymmetric $Z_2 \times Z_2$
amplitudes, when brane supersymmetry breaking is present 
additional characters are
needed to describe the open strings stretched between branes and
antibranes. These new characters, that we denote by $\tau^{(-)}_{kl}$, are 
obtained from the others in eq. (\ref{a3}) interchanging 
$O_2$ with $V_2$ and $S_2$ with $C_2$ in the last three factors, that 
correspond to the three internal tori:
\ba
\t^{(-)}_{oo} &=& O_2O_2O_2O_2+V_2V_2V_2V_2-C_2S_2S_2S_2-S_2C_2C_2C_2 \ , 
\nonumber 
\\
\t^{(-)}_{og} &=& V_2V_2O_2O_2+O_2O_2V_2V_2-S_2C_2S_2S_2-C_2S_2C_2C_2 \ , 
\nonumber 
\\
\t^{(-)}_{oh} &=& V_2O_2O_2V_2+O_2V_2V_2O_2-S_2S_2S_2C_2-C_2C_2C_2S_2 \ , 
\nonumber 
\\
\t^{(-)}_{of} &=& V_2O_2V_2O_2+O_2V_2O_2V_2-S_2S_2C_2S_2-C_2C_2S_2C_2 \ , 
\nonumber 
\\
\t^{(-)}_{go} &=& O_2O_2S_2C_2+V_2V_2C_2S_2-C_2S_2V_2O_2-S_2C_2O_2V_2 \ , 
\nonumber 
\\
\t^{(-)}_{gg} &=& V_2V_2S_2C_2+O_2O_2C_2S_2-C_2S_2O_2V_2-S_2C_2V_2O_2 \ , 
\nonumber 
\\
\t^{(-)}_{gh} &=& V_2O_2S_2S_2+O_2V_2C_2C_2-S_2S_2V_2V_2-C_2C_2O_2O_2 \ , 
\nonumber 
\\
\t^{(-)}_{gf} &=& V_2O_2C_2C_2+O_2V_2S_2S_2-C_2C_2V_2V_2-S_2S_2O_2O_2 \ , 
\nonumber 
\\
\t^{(-)}_{ho} &=& O_2S_2C_2O_2+V_2C_2S_2V_2-S_2O_2V_2C_2-C_2V_2O_2S_2 \ , 
\nonumber 
\\
\t^{(-)}_{hg} &=& V_2C_2C_2O_2+O_2S_2S_2V_2-S_2O_2O_2S_2-C_2V_2V_2C_2 \ , 
\nonumber 
\\
\t^{(-)}_{hh} &=& V_2S_2C_2V_2+O_2C_2S_2O_2-C_2O_2V_2S_2-S_2V_2O_2C_2 \ , 
\nonumber 
\\
\t^{(-)}_{hf} &=& V_2S_2S_2O_2+O_2C_2C_2V_2-S_2V_2V_2S_2-C_2O_2O_2C_2 \ , 
\nonumber 
\\
\t^{(-)}_{fo} &=& O_2S_2O_2C_2+V_2C_2V_2S_2-C_2V_2S_2O_2-S_2O_2C_2V_2 \ , 
\nonumber 
\\
\t^{(-)}_{fg} &=& V_2C_2O_2C_2+O_2S_2V_2S_2-S_2O_2S_2O_2-C_2V_2C_2V_2 \ , 
\nonumber 
\\
\t^{(-)}_{fh} &=& V_2S_2O_2S_2+O_2C_2V_2C_2-S_2V_2S_2V_2-C_2O_2C_2O_2 \ , 
\nonumber 
\\
\t^{(-)}_{ff} &=& V_2S_2V_2C_2+O_2C_2O_2S_2-S_2V_2C_2O_2-C_2O_2S_2V_2 \ . 
\ea

\section{Characters for the $T^6/Z_4$ orbifold}

In this Appendix we list the characters needed for the
$Z_4$ orientifold of Section 3:
\ba
\rho_{00} &=& V_2O_2\chi_0\xi_0+O_2V_2\chi_0\xi_4-S_2C_2\chi_0\xi_2-C_2S_2
\chi_0\xi_{-2} \ , 
\nonumber 
\\
\rho_{01} &=& V_2V_2\chi_{1/2}\xi_{-2}+O_2O_2\chi_{1/2}\xi_2-S_2S_2\chi_{1/2}
\xi_0-C_2C_2\chi_{1/2}\xi_{4} \ , 
\nonumber 
\\
\rho_{02} &=& V_2O_2\chi_0\xi_4+O_2V_2\chi_0\xi_0-S_2C_2\chi_0\xi_{-2}
-C_2S_2\chi_0\xi_{2} \ , 
\nonumber 
\\
\rho_{03} &=& V_2V_2\chi_{1/2}\xi_2+ O_2O_2\chi_{1/2}\xi_{-2}
-S_2S_2\chi_{1/2}\xi_4-C_2C_2\chi_{1/2}\xi_{0} \ , 
\nonumber 
\\
\rho_{10} &=& V_2S_2\chi_{0}\xi_{-3} + O_2C_2\chi_{0}\xi_{1}
-S_2O_2\chi_{0}\xi_{-1}-C_2V_2\chi_{0}\xi_{3} \ , 
\nonumber 
\\
\rho_{11} &=& V_2C_2\chi_{1/2}\xi_{3}+ O_2S_2\chi_{1/2}\xi_{-1}
-S_2V_2\chi_{1/2}\xi_{-3} -C_2O_2\chi_{1/2}\xi_{1} \ , 
\nonumber 
\\
\rho_{12} &=& V_2S_2\chi_{0}\xi_{1}+ O_2C_2\chi_{0}\xi_{-3}
-S_2O_2\chi_{0}\xi_{3}-C_2V_2\chi_{0}\xi_{-1} \ , 
\nonumber 
\\
\rho_{13} &=& V_2C_2\chi_{1/2}\xi_{-1}+ O_2S_2\chi_{1/2}\xi_{3}
-S_2V_2\chi_{1/2}\xi_{1} -C_2O_2\chi_{1/2}\xi_{-3} \ , 
\nonumber 
\\
\rho_{20} &=& V_2O_2\chi_{1/2}\xi_0+O_2V_2\chi_{1/2}\xi_4
-S_2C_2\chi_{1/2}\xi_2-C_2S_2\chi_{1/2}\xi_{-2} \ , 
\nonumber 
\\
\rho_{21} &=& V_2V_2\chi_{0}\xi_{-2}+O_2O_2\chi_{0}\xi_2-S_2S_2\chi_{0}
\xi_0-C_2C_2\chi_{0}\xi_{4} \ , 
\nonumber 
\\
\rho_{22} &=& V_2O_2\chi_{1/2}\xi_4+O_2V_2\chi_{1/2}\xi_0
-S_2C_2\chi_{1/2}\xi_{-2}-C_2S_2\chi_{1/2}\xi_{2} \ , 
\nonumber 
\\
\rho_{23} &=& V_2V_2\chi_{0}\xi_2+O_2O_2\chi_{0}\xi_{-2}
-S_2S_2\chi_{0}\xi_4-C_2C_2\chi_{0}\xi_{0} \ ,
\nonumber
\\
\rho_{30} &=& V_2C_2\chi_{0}\xi_{3}+ O_2S_2\chi_{0}\xi_{-1}
-S_2V_2\chi_{0}\xi_{-3} -C_2O_2\chi_{0}\xi_{1} \ , 
\nonumber 
\\
\rho_{31} &=& V_2S_2\chi_{1/2}\xi_{1}+ O_2C_2\chi_{1/2}\xi_{-3}
-S_2O_2\chi_{1/2}\xi_{3}-C_2V_2\chi_{1/2}\xi_{-1} \ , 
\nonumber 
\\
\rho_{32} &=& V_2C_2\chi_{0}\xi_{-1}+ O_2S_2\chi_{0}\xi_{3}
-S_2V_2\chi_{0}\xi_{1} -C_2O_2\chi_{0}\xi_{-3} \ , 
\nonumber 
\\
\rho_{33} &=& V_2S_2\chi_{1/2}\xi_{-3}+ O_2C_2\chi_{1/2}\xi_{1}
-S_2O_2\chi_{1/2}\xi_{-1}-C_2V_2\chi_{1/2}\xi_{3} \ .
\ea
The $\s$ characters are obtained interchanging in the internal part
$O_2$ with $V_2$, $\xi_0$ with $\xi_4$, $S_2$ with $C_2$, 
and $\xi_2$ with $\xi_{-2}$:
\ba
\s_{00} &=& V_2 V_2 \chi_0 \xi_4 + O_2 O_2 \chi_0 \xi_0 -
S_2 S_2 \chi_0 \xi_{-2} - C_2 C_2 \chi_0\xi_{2} \ ,
\nonumber  
\\
\s_{01} &=& V_2 O_2 \chi_{1/2} \xi_{2} + O_2 V_2 \chi_{1/2} \xi_{-2} 
- S_2 C_2 \chi_{1/2} \xi_4 - C_2 S_2 \chi_{1/2} \xi_{0} \ ,
\nonumber
\\
\s_{02} &=& V_2 V_2 \chi_0 \xi_0 + O_2 O_2 \chi_0 \xi_4
- S_2 S_2 \chi_0 \xi_{2} - C_2 C_2 \chi_0 \xi_{-2} \ ,
\nonumber
\\
\s_{03} &=& V_2 O_2 \chi_{1/2} \xi_{-2} + O_2 V_2 \chi_{1/2} \xi_{2}  
- S_2 C_2 \chi_{1/2} \xi_0 - C_2 S_2 \chi_{1/2} \xi_{4} \ , \nonumber       
\\
\s_{20} &=& O_2O_2\chi_{1/2}\xi_{0} + V_2V_2\chi_{1/2}\xi_{4}
- S_2S_2\chi_{1/2}\xi_{-2} - C_2C_2\chi_{1/2}\xi_{2} \ , 
\nonumber
\\
\s_{21} &=& V_2O_2\chi_{0}\xi_{-2} + O_2V_2\chi_{0}\xi_{2} 
- S_2C_2\chi_{0}\xi_{0} - C_2S_2\chi_{0}\xi_{4} \ ,
\nonumber 
\\
\s_{22} &=& V_2V_2\chi_{1/2}\xi_{0} + O_2O_2\chi_{1/2}\xi_{4} 
- S_2S_2\chi_{1/2}\xi_{2} - C_2C_2\chi_{1/2}\xi_{-2} \ , 
\nonumber
\\
\s_{23} &=& V_2O_2\chi_{0}\xi_{2} + O_2V_2\chi_{0}\xi_{-2} 
- S_2C_2\chi_{0}\xi_{4} - C_2S_2\chi_{0}\xi_{0} \ .
\ea

\vfill\eject


\begin{thebibliography}{99}
\bibitem{carg}{A. Sagnotti, in: Cargese '87, Non-Perturbative Quantum 
Field Theory, eds. G. Mack et al. (Pergamon Press, Oxford, 1988) p. 521;
G. Pradisi and A. Sagnotti, \PLB{216}{89}{59};
M. Bianchi and A. Sagnotti, \PLB{247}{90}{517}, \NPB{361}{91}{519}.}
\bibitem{a} I. Antoniadis, \PLB{246}{90}{377}.
\bibitem{w} E. Witten, \NPB{471}{96}{135}; J.D. Lykken, \PRD{54}{96}{3693}.
\bibitem{add} N. Arkani-Hamed, S. Dimopoulos and G. Dvali, \PLB{429}{98}{263};
K.R. Dienes, E. Dudas and T. Gherghetta, \PLB{436}{98}{55};
I. Antoniadis, N. Arkani-Hamed, S. Dimopoulos and G. Dvali,
\PLB{436}{98}{263}; for a recent review see I. Antoniadis, hep-th/9909212
and references therein.
\bibitem{int} K. Benakli, \PRD{60}{99}{104002}; 
C. Burgess, L.E. Ib{\'a}{\~n}ez and F. Quevedo, \PLB{447}{99}{257}.
\bibitem{pc}J. Polchinski and Y.C. Cai, \NPB{296}{88}{91}. 
\bibitem{pol}J. Polchinski, \PRL{75}{95}{4724}.
\bibitem{erice} M. Bianchi, Ph.D. Thesis, preprint ROM2F-92/13;
A. Sagnotti, hep-th/9302099.
\bibitem{zw} G. Zwart, \NPB{526}{98}{378}; 
Z. Kakushadze, G. Shiu and S.H.H. Tye, \NPB{533}{98}{25}; 
G. Aldazabal, A. Font, L.E. Ib{\'a}{\~n}ez and G. Violero, \NPB{536}{98}{29}.
\bibitem{ads2} I. Antoniadis, E. Dudas and A. Sagnotti, \PLB{464}{99}{38}.
\bibitem{closed}R. Rohm, \NPB{237}{84}{553};
C. Kounnas and M. Porrati, \NPB{310}{88}{355}; 
I. Antoniadis, C. Bachas, D.C. Lewellen and T.N. Tomaras, \PLB{207}{88}{441};
S. Ferrara, C. Kounnas, M. Porrati and F. Zwirner, \NPB{318}{89}{75};
C. Kounnas and B. Rostand, \NPB{341}{90}{641};
I. Antoniadis, \PLB{246}{90}{377};
I. Antoniadis and C. Kounnas, \PLB{261}{91}{369}.
\bibitem{ads} I. Antoniadis, E. Dudas and A. Sagnotti,
\NPB{544}{99}{469};
I. Antoniadis, G. D'Appollonio, E. Dudas and
A. Sagnotti, \NPB{553}{99}{133}.
\bibitem{aaf} R. Blumenhagen and L. G{\"o}rlich, \NPB{551}{99}{601}; 
C. Angelantonj, I. Antoniadis and K. F{\"o}rger, \NPB{555}{99}{116}.
\bibitem{adds2} I. Antoniadis, G. D'Appollonio, E. Dudas and
A. Sagnotti, hep-th/9907184, to appear in {\sl Nucl. Phys.} {\bf B}; 
A.L. Cotrone, hep-th/9909116.
\bibitem{bachas} C. Bachas, hep-th/9503030; M. Bianchi and Ya.S. Stanev,
\NPB{253}{98}{193}. 
\bibitem{bmp}M. Bianchi, J.F. Morales and G. Pradisi, hep-th/9910228. 
\bibitem{su} S. Sugimoto, hep-th/9905159.
\bibitem{au} G. Aldazabal and A.M. Uranga, hep-th/9908072.
\bibitem{abpss} C.~Angelantonj, M.~Bianchi, G.~Pradisi, A.~Sagnotti and 
Ya.S.~Stanev, \PLB{385}{96}{96}.
\bibitem{aiq} G. Aldazabal, L.E. Ib{\'a}{\~n}ez and F. Quevedo, hep-th/9909172.
\bibitem{sen} A. Sen, \JHEP{9806}{98}{007}, {\bf 9808} (1998) 010, 012,
{\bf 9809} (1998) 023, {\bf 9812} (1998) 021. For recent reviews, see 
A. Sen, hep-th/9904207; A. Lerda and R. Russo, hep-th/9905006.
\bibitem{vafa} C. Vafa, \NPB{273}{86}{592}; C. Vafa and E. Witten,
{\it J. Geom. Phys.} {\bf 15} (1995) 189; J. Blum, \NPB{486}{97}{34};
M.R. Douglas, hep-th/9807235.
\bibitem{pss} D. Fioravanti, G. Pradisi and A. Sagnotti, \PLB{321}{94}{349};
G. Pradisi, A. Sagnotti and Ya.S. Stanev, \PLB{354}{95}{279},
{\bf B356} (1995) 230, {\bf B381} (1996) 97.
\bibitem{bl} M. Berkooz and R.G. Leigh, \NPB{483}{97}{187}.
\bibitem{kakul}Z. Kakushadze, G. Shiu and S.H.H. Tye, in \cite{zw};
K. Mohri, \NPB{521}{98}{161}; S. Mukhopadhyay and K. Ray, hep-th/9909107.
\bibitem{toroidal} M. Bianchi, G. Pradisi and A. Sagnotti, \NPB{376}{92}{365}
\bibitem{gs} M.B. Green and J.H. Schwarz, \PLB{149}{84}{117}.
\bibitem{comm} Z. Kakushadze, G. Shiu and S.H.H. Tye, \PRD{58}{98}{086001};
C. Angelantonj, hep-th/9908064, to appear in {\it Nucl. Phys.} {\bf B}.
\bibitem{notach} A. Sagnotti, hep-th/9509080, hep-th/9702093, 
C. Angelantonj, \PLB{444}{98}{309};
R. Blumenhagen, A. Font and D. L{\"u}st, hep-th/9904069;
R. Blumenhagen and A. Kumar, hep-th/9906234; 
K. F{\"o}rger, hep-th/9909010.
\bibitem{bs} M. Bianchi and A. Sagnotti, in \cite{carg}.
\bibitem{gp} E. Gimon and J. Polchinski, hep-th/9601038.
\bibitem{ggs} A. Sagnotti, \PLB{294}{92}{196}.
\bibitem{bd} J. Blum and K.R. Dienes, \PLB{414}{97}{260},
\NPB{516}{98}{83}.
\bibitem{bkl} R. Blumenhagen, C. Kounnas and D. L{\"u}st, hep-th/9910094.
\bibitem{aq} I. Antoniadis and M. Quiros, \PLB{392}{97}{61};
E. Dudas and C. Grojean, \NPB{507}{97}{553}, hep-th/9704177; I. Antoniadis and
M. Quiros, \NPB{505}{97}{109}, hep-th/9705037, \PLB{416}{98}{327};
E. Dudas, \PLB{416}{98}{309}.
\end{thebibliography}
\end{document}